\documentclass[final, 5p, twocolumn,sort&compress]{elsarticle}

\makeatletter
\def\ps@pprintTitle{%
  \let\@oddhead\@empty
  \let\@evenhead\@empty
  \def\@oddfoot{\reset@font\hfil}%
  \let\@evenfoot\@oddfoot
}
\makeatother

\usepackage{amssymb}
\usepackage{amsmath}

\usepackage{float}            
\usepackage{caption}          
\usepackage{subcaption}       
\usepackage{booktabs}         

\usepackage[table]{xcolor}    
\definecolor{coltau}{HTML}{dbb40c} 
\definecolor{colmu}{HTML}{7e1e9c}  
\definecolor{cole}{HTML}{f97306}   

\usepackage{hyperref}        
\hypersetup{
    breaklinks=true,
    colorlinks=true,
    linkcolor=blue,
    citecolor=blue,
    urlcolor=blue
}

\usepackage{layouts}          
\usepackage{verbatim}

\newcommand{\be}{\begin{equation}}
\newcommand{\ee}{\end{equation}}
\newcommand{\bea}{\begin{eqnarray}}
\newcommand{\eea}{\end{eqnarray}}

\newcommand{\lp}{\left(}
\newcommand{\rp}{\right)}

\newcommand{\dNeff}{\Delta N_{\rm eff}}
\newcommand{\Ht}{\tilde{H}}
\newcommand{\gt}{\tilde{g}}

\newcommand{\CLASS}{\texttt{CLASS}} 

\journal{Physics of the Dark Universe}

\begin{document}

\begin{frontmatter}

\title{\textbf{Improved cosmological constraints on axion-lepton interactions}}

\author{Marcin Badziak}
\ead{Marcin.Badziak@fuw.edu.pl}
\author{Adam Gomułka\corref{cor1}}
\cortext[cor1]{Corresponding author}
\ead{Adam.Gomulka@fuw.edu.pl}
\author{Maxim Laletin}
\ead{Maxim.Laletin@fuw.edu.pl}
\author{Krzysztof Szafrański}
\ead{Krzysztof.Szafranski@fuw.edu.pl}

\address{{Institute of Theoretical Physics, Faculty of Physics, University of Warsaw},
            {ul. Pasteura 5},
            {Warsaw},
            {PL-02-093},
            {Poland}}

\begin{abstract}
We present updated cosmological constraints on axion-lepton interactions based on state-of-the-art computations of the thermal axion abundance. By combining Planck Cosmic Microwave Background (CMB) data with baryon acoustic oscillation (BAO) measurements from DESI DR2, we derive improved limits on both lepton-flavor-conserving (LFC) and lepton-flavor-violating (LFV) axion couplings. Incorporating finite axion mass effects substantially strengthens the bounds for axion masses above 0.1~eV compared to those inferred from the $\Delta N_{\rm eff}$ constraint alone. The bounds on the LFC axion-tau coupling and LFV axion couplings to tau and muon or electron are improved by several orders of magnitude and 
the lower bound on the axion decay constant may exceed $10^6$ and $10^8$~GeV, respectively, for axion masses above 1~eV. Our cosmological constraints on LFC axion couplings to muons and taus and LFV axion couplings to tau and muon or electron are stronger than all other constraints for masses above 0.3~eV. In particular, they are stronger than recent collider constraints from Belle-II on $\tau\to la$ decays, where $l=e$ or $\mu$. The collider constraints on $\mu\to ea$ decays are weaker than the cosmological constraints for axion masses above 100~eV. Our results are relevant for both the QCD axion and axion-like particles (ALPs).
\end{abstract}

\end{frontmatter}

\section{Introduction}
\label{sec1}
{\let\thefootnote\relax\footnotetext{\copyright~2026. This manuscript version is made available under the CC BY-NC-ND 4.0 license (\url{http://creativecommons.org/licenses/by-nc-nd/4.0/}). The final authenticated version is available online at: \url{https://doi.org/10.1016/j.dark.2026.102335}}}
Pseudo-Nambu-Goldstone bosons (PNGB) arising from spontaneously broken approximate global symmetries appear in many well-motivated extensions of the Standard Model (SM). A particularly well-motivated PNGB is the QCD axion which arises from spontaneously broken Peccei-Quinn (PQ) symmetry which is introduced to solve the strong CP problem~\citep{Peccei:1977hh}. The QCD axion is a good dark matter (DM) candidate~\citep{Preskill:1982cy,Abbott:1982af,Dine:1982ah} and can be linked to a mechanism responsible for the observed baryon asymmetry of the Universe~\citep{Co:2019wyp}. PQ symmetry is broken only by non-perturbative QCD effects leading to a small but non-zero axion mass which is entirely determined by the axion decay constant $f$~\citep{GrillidiCortona:2015jxo}:
\begin{align}
    m_a \approx0.569(51) \left(\frac{10^7 \text{ GeV}}{f}\right) \text{ eV}.
    \label{eq: axion zero temperature mass}
\end{align}
In QCD axion models there is a lower bound on $f$ stemming from astrophysical constraints. In minimal QCD axion models, such as KSVZ~\citep{Kim:1979if,Shifman:1979if} or DFSZ~\citep{Dine:1981rt,Zhitnitsky:1980tq} models, the lower bound on the QCD axion decay constant is at the level of few times $10^8$~GeV which translates to the QCD axion mass of order $\mathcal{O}(0.01)$~eV~\citep{Buschmann:2021juv,Carenza:2020cis}. In the so-called astrophobic axion models~\citep{DiLuzio:2017ogq,Bjorkeroth:2019jtx,Badziak:2021apn,DiLuzio:2022tyc,Takahashi:2023vhv,Badziak:2023fsc,Badziak:2024szg} this bound can be relaxed and a value of $f$ even as small as few times $10^6$~GeV~\citep{Badziak:2023fsc} may be allowed leading to the QCD axion mass $\mathcal{O}(1)$~eV. One should, however, keep in mind that the astrophysical constraints rely on phenomenological models of stars which are not fully understood, thus, they suffer from large theoretical uncertainties.

The above mentioned astrophysical bounds and the axion mass prediction~\eqref{eq: axion zero temperature mass} are specific for the QCD axion. For general PNGB, also known as axion-like particle, that we refer to simply as axion, $f$ and $m_a$ are independent parameters and, in particular, the axion mass can span many orders of magnitude and can exceed eV. 

Axions are very good candidates for cold dark matter (CDM) since they are produced via the misalignment mechanism~\citep{Preskill:1982cy,Abbott:1982af,Dine:1982ah,Co:2019jts} or decay of topological defects~\citep{Vilenkin:1982ks,Kawasaki:2014sqa,Buschmann:2019icd,Gorghetto:2020qws}.
However, axions can be also produced thermally via interactions with SM particles in the early Universe plasma. In particular, thermally produced axions can play the role of the observed DM~\citep{Panci:2022wlc,Aghaie:2024jkj,DEramo:2025jsb} provided that their lifetime is long enough 
and their mass is above the warm dark matter (wDM) bound from Ly-$\alpha$ forests~\citep{Boyarsky:2008xj}, which have been shown to be $\mathcal{O}(10)$~keV for axions~\citep{Kamada:2019kpe,Ballesteros:2020adh,DEramo:2020gpr,Decant:2021mhj,DEramo:2025jsb}. 

For smaller masses, axions are not cold enough to explain all the observed DM, but may leave imprints in various cosmological observables, so the cosmological data allow to set constraints on the axion decay constant which are more robust than the astrophysical ones. Most studies so far have focused on very light, ultra-relativistic axions which contribute to dark radiation and increase the additional effective number of neutrinos $\dNeff$, which is constrained by the Big Bang Nucleosynthesis (BBN) and the Cosmic Microwave Background (CMB). In particular, the Planck satellite measurements of CMB anisotropies set an upper bound on $\dNeff\lesssim0.3$ at 95\% credible level (C.L.)~\citep{Planck:2018vyg} which leads to an upper bound on axion couplings. There have been numerous papers devoted to the computation of axion contribution to $\dNeff$, see e.g. Refs.~\citep{Berezhiani:1992rk,Chang:1993gm,Hannestad:2005df,Salvio:2013iaa,Ferreira:2018vjj,DEramo:2018vss,Arias-Aragon:2020shv,Ferreira:2020bpb,Ghosh:2020vti,Green:2021hjh,DEramo:2021usm,DEramo:2021psx,DEramo:2021lgb,Langhoff:2022bij,Notari:2022ffe,Bianchini:2023ubu,Badziak:2024szg,Dunsky:2022uoq,Bouzoud:2024bom,Badziak:2024qjg,DEramo:2024jhn}.
The majority of works computed $\dNeff$ using instantaneous decoupling approximation or by solving Boltzmann equations for the axion number density assuming that they are in thermal equilibrium. It was recently realised that in order to obtain precise predictions for $\dNeff$ that fully exploits the expected sensitivity of planned CMB experiments, such as Simons Observatory~\citep{SimonsObservatory:2018koc}, one has to relax the assumption of thermal distribution functions and solve Boltzmann equations for the actual distribution function. This improved approach was used to set constraints on the KSVZ model of the QCD axion~\citep{Notari:2022ffe,Bianchini:2023ubu} and model-independent constraints on axion couplings to SM fermions~\citep{Badziak:2024qjg,DEramo:2024jhn}. 

It is important to emphasize, however, that CMB constraints on $\dNeff$ can be used to obtain reliable constraints on axion couplings only for axion masses below $\mathcal{O}(0.1)$~eV. For larger masses axions are no longer ultra-relativistic during recombination and a dedicated cosmological analysis is required to derive constraints on axion couplings. Since the axion mass gives additional contribution to the axion energy density during and after recombination the constraints for massive axions are generically stronger than those derived solely from $\dNeff$. This have been explicitly shown for 
DFSZ~\citep{Ferreira:2020bpb,DEramo:2022nvb} and KSVZ~\citep{DEramo:2022nvb,Notari:2022ffe,Bianchini:2023ubu} models for the QCD axion. To the best of our knowledge, model-independent cosmological constraints including axion mass have been derived only for the axion-photon and axion-gluon couplings~\citep{Caloni:2022uya} (but with the axion energy density obtained under instantaneous decoupling approximation). 

The main objective of the present paper is to derive model-independent cosmological constraints on axion couplings to leptons for a large range of axion masses including those larger than the recombination temperature. Constraints on axion-lepton couplings have been studied before using $\dNeff$ in Ref.~\citep{Ghosh:2020vti}, but as emphasised above, they underestimate constraints for axion masses above $\mathcal{O}(0.1)$~eV. We use CMB data from Planck~\citep{Planck:2019nip, Planck:2018lbu} and Baryon Acoustic Oscillation (BAO) data from the DESI DR2 release~\citep{DESI:2025zgx}.
We consider both flavor-conserving and flavor-violating axion couplings. The former couplings are present in a large class of axion models, including the DFSZ model, while the latter couplings arise generically in models where the PQ symmetry acts as a flavor symmetry explaining the observed hierarchy of the SM fermion masses~\cite{Davidson:1981zd,Wilczek:1982rv,Berezhiani:1989fp,Ema:2016ops,Calibbi:2016hwq} and in minimal astrophobic models of the QCD axion~\cite{DiLuzio:2017ogq,Badziak:2021apn,Badziak:2024szg}. There are two other important improvements in our analysis that make the derived constraints more robust than those in the existing literature. First: we use the state-of-the-art computations of thermal axion abundance based on solving full Boltzmann equations for the momentum distribution functions, as recently done in Ref.~\citep{Badziak:2024qjg}. This allows for a precise determination of axion energy density both in the case of axions in thermal equilibrium and in the case of axions produced via freeze-in~\citep{Hall:2009bx}. Second: we perform a full cosmological analysis with these correct non-thermal axion distributions, instead of relying on a point-estimate to set our constraints. 

We demonstrate that the effect of axion mass is very significant and may improve the constraints on axion couplings even by several orders of magnitude. In particular, we show that for axion masses above $\mathcal{O}(0.1)$~eV cosmology provides the most stringent constraints on axion couplings to muons and taus as well as on flavor-violating axion couplings to $\tau e$ and $\tau\mu$.

The rest of the article is organized as follows. In Section~\ref{Sec:axion_production} we describe the methodology behind the computation of the thermal axion abundance and show the results for axion energy density for various production channels induced by axion-lepton interactions. In Section~\ref{sec:methodology} we introduce the methodology of our cosmological analysis. In Section~\ref{sec:results} we demonstrate and analyze the resulting constraints on the axion-lepton couplings for several fixed values of the axion mass. We conclude in Section~\ref{sec:conclusions}. Finally, in a series of Appendices we present some technical details of our cosmological analysis as well as constraints on the axion parameter space when both axion couplings and axion mass are varied simultaneously. We also elaborate on how the existence of axions impacts the CMB temperature and matter power spectra for various axion masses.

\section{Thermal production of axions}
\label{Sec:axion_production}

In this work we consider theories which include an axion field $a$ and are described by the following effective Lagrangian:
\begin{equation}
    \mathcal{L}= \mathcal{L}_{\rm SM} +\frac{1}{2}(\partial_\mu a)^2 -\frac{1}{2}m_a^2a^2 + \mathcal{L}^{(a)}_{\rm int}  \,,
\end{equation}
where $\mathcal{L}_{\rm SM}$ is the SM Lagrangian, $m_a$ is the axion mass and $\mathcal{L}^{(a)}_{\rm int}$ includes axion interactions with SM fields that we specify later. The axion field can be thought of as a pseudo-Nambu-Goldstone boson (PNGB) of some spontaneously broken global symmetry. The explicit breaking of a global symmetry is parameterised by the axion mass. In this work we take $m_a$ to be a free parameter, but we will also discuss a special case of the QCD axion predicted by models that solve the strong CP problem via PQ mechanism where $m_a$ is generated solely by non-perturbative QCD effects.

In the present work we focus on axion interactions with leptons which are given by:
\begin{equation}
  \mathcal{L}^{(a)}_{\rm int} =  \frac{\partial_\mu a}{2 f} \; \overline{l}_i \gamma^\mu \left( C^V_{ij} + C^A_{ij} \gamma_5 \right) l_j \,,
\end{equation}
where $f$ is the axion decay constant. Note that axion interactions with leptons do not necessarily conserve flavor. For further convenience we define $C_{i}\equiv C^A_{ii}$ and $C_{ij}\equiv \sqrt{|C_{ij}^A|^2 + |C_{ij}^V|^2}$. We will refer collectively to flavor-conserving and flavor-violating couplings as $C$. 

In order to derive the cosmological bounds we need to know the abundance of axions produced in the SM bath. We calculate it following the same methodology as in Ref.~\cite{Badziak:2024qjg}, namely by solving the Boltzmann equation for the axion phase-space distribution function $f_a (x,q)$ in the expanding Friedmann–Lemaître–Robertson–Walker (FLRW) Universe\footnote{This momentum-dependent approach is also described in detail for two-body annihilations and elastic scatterings of massive particles in Ref.~\cite{Binder:2017rgn} and for dark radiation produced from two-body decays and binary scatterings in Ref.~\cite{DEramo:2023nzt}.}

\be
\Ht (x) \, \left( x\partial_x - \gt q\partial_q \right) f_a(x,q) = \mathcal{C}[f_a] \, ,
\ee
where $x = m/T$ is the effective unit of cosmic time inversely related to the temperature of the SM bath ($m$ can be arbitrary, but it is convenient to use the mass of the heaviest lepton in the reaction), $q = p/T$ is the reduced momentum of the axion, $\Ht = H/(1 + \gt)$ is the reduced Hubble parameter with $\gt = (-1/3) (d \ln h_s / d \ln x)$ being the derivative of the effective number of entropy degrees of freedom $h_s$ and $\mathcal{C}[f_a]$ is the collision term that accounts for the impact of all the processes involving an axion on its distribution function. The exact formulae for the collision terms and interaction rates, as well as other practical details can be found e.g. in Ref.~\cite{Badziak:2024qjg}. The axion abundance is determined by the integration of $f_a (x,q)$ over the phase-space of the axion, be it the number density of axions $n_a$ or their energy density $\rho_a$. As demonstrated in Ref.~\cite{Badziak:2024qjg} the axion abundance computed by solving the Boltzmann equation for the number density $n_a$ under the assumption of kinetic equilibrium can noticeably deviate from the one recovered with the use of the more general momentum-dependent Boltzmann equation for the distribution function $f_a (x,q)$. However, the benefits of the latter approach extend beyond just a more precise estimate of the axion abundance -- below we discuss the importance of axion distribution function $f_a (x,q)$ for cosmological probes. Throughout the text, whenever the distinction between the two approaches to compute the axion abundance is discussed, the first will be referred to as fBE (full phase-space Boltzmann equation) and the second as nBE (number-density Boltzmann equation).

In the presence of flavor-conserving axion-lepton couplings, axions are produced through scattering processes involving leptons and photons: $l^\pm \gamma \to l^\pm a$, $l^+ l^- \to \gamma a$. The corresponding scattering rates are proportional to $|C_i|^2$. For flavor-violating models, axion production is dominated  by the decays of the form $l_i^\pm \to l_j^\pm a$ with the corresponding decay rates proportional to $|C_{ij}|^2$. Flavor-violating scatterings involving photons are subdominant~\cite{Aghaie:2024jkj} and can be safely neglected when deriving cosmological constraints. 

\begin{figure}
    \centering
    \includegraphics[width=\columnwidth]{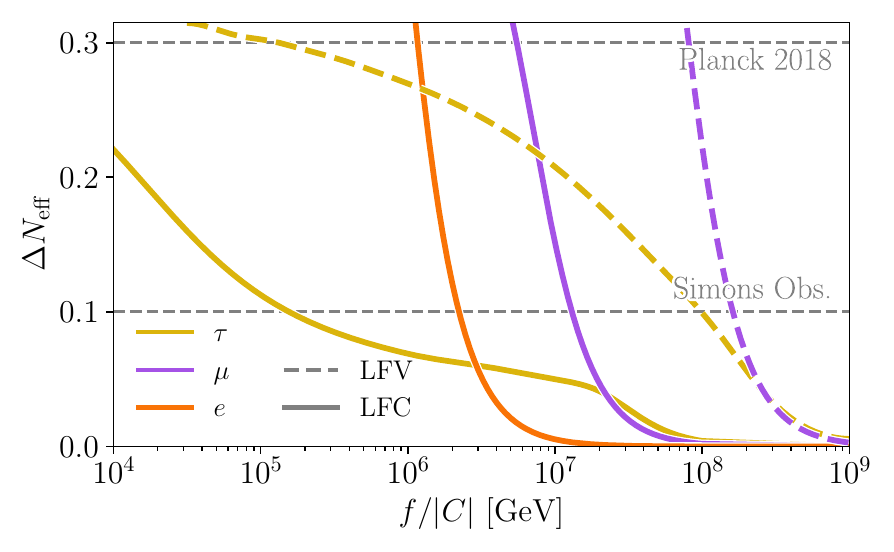}
    \caption{$\dNeff$ computed via fBE as a function of $f/|C|$ for all considered lepton channels. Solid curves correspond to lepton-flavor conserving (LFC) processes, while dashed curves correspond to lepton-flavor violating (LFV) processes. The results for electron interactions are plotted with an orange curve (only LFC), muon interactions -- with purple curves, and tau interactions -- with golden curves. Dashed grey horizontal lines mark the current constraint from Planck and the predicted sensitivity of Simons Observatory. See text for more details.
    }
    \label{fig:dNeff_all}
\end{figure}

Cosmological constraints for axions are often derived in terms of the effective number of additional neutrino species, which is given by the following formula for axions

\begin{equation}
    \dNeff = \frac{8}{7} \lp \frac{11}{4} \rp^{4/3} \frac{\rho_a}{\rho_{\gamma}} \, ,
    \label{eq: dNeff}
\end{equation}
where $\rho_a$ and $\rho_{\gamma}$ are the axion energy density and photon energy density, respectively. $\dNeff$ effectively parameterizes the impact of additional relativistic species on the sound horizon around recombination, fluctuation amplitude and other cosmological quantities, thus precision CMB measurements and BAO measurements can directly constrain the presence of any excessive radiation density above the SM. For example, the benchmark constraint value established by the Planck analysis is $\dNeff \leq 0.3$ at 95\% C.L. \cite{Planck:2018vyg}.
In Fig.~\ref{fig:dNeff_all} we show $\dNeff$ computed via fBE approach for all axion-lepton couplings that we consider in this work and show the corresponding constraints from Planck and the projected sensitivity of Simons Observatory.

When axion contribution to $\dNeff$ is computed it is often assumed that axions were produced in thermal equilibrium and remain ultra-relativistic at the onset of recombination. As demonstrated in Ref.~\cite{Badziak:2024qjg} the former assumption (being expressed as the relation between axion number density and energy density) can significantly underestimate the energy density of axions if axion production is rather mild (freeze-in scenario). Furthermore, if axion mass is large enough, the non-relativistic axion species start to dominate the energy density of axions around recombination 
\be
\rho_a = \frac{g_a T^4}{2\pi^2} \int dq \, \sqrt{q^2 + \left(\frac{m_a}{T}\right)^2} \cdot q^2 f_a(q)
\ee
and $\dNeff$ can no longer be used to obtain reliable constraint on the axion decay constant.
One can think of a modification to Eq.~\eqref{eq: dNeff} which would take into account only the energy density of \textit{relativistic} axions at the given temperature. However, the non-relativistic component of the axion gas starts to play its part in the structure formation and leave an imprint on several cosmological observables too. Hence, imposing the constraints on axion models gets more involved than computing just one effective parameter. Still, one may expect that the inclusion of the axion mass makes the constraint stronger than that derived from $\dNeff$ for a would-be massless axion, because the axion energy density is enhanced if the axion mass is above the recombination temperature.  
Indeed, this effect was observed in several detailed analyses of cosmological constraints on QCD axion models, such as KSVZ~\cite{DEramo:2022nvb,Notari:2022ffe,Bianchini:2023ubu} and DFSZ~\cite{DEramo:2022nvb}. Model-independent cosmological constraints including axion mass were also derived on the axion-photon and axion-gluon couplings~\cite{Caloni:2022uya}. However, model-independent cosmological constraints on axion couplings to leptons that take into account the axion mass are missing and our analysis feels this gap.

Moreover, the cosmological effects of the axions also become sensitive to the shape of the momentum distribution $f_a (x,q)$ as it defines the proportions of non-relativistic and relativistic axions. In order to derive cosmological constraints on axion-lepton couplings we do not assume thermal momentum distribution functions, but use those derived from fBE. The cosmological analyses that use non-thermal distributions functions have been performed only for the KSVZ model~\cite{Notari:2022ffe,Bianchini:2023ubu}.

\section{Cosmological analysis methodology}
\label{sec:methodology}

We consider a cosmological model that extends the standard $\Lambda$CDM by incorporating both massive axions and massive neutrinos. The baseline $\Lambda$CDM model is typically described by six parameters: the baryon and CDM energy densities, $\omega_b$ and $\omega_{\rm cdm}$~\footnote{Note that we do not try to explain the observed relic abundance of DM with axions, so $\omega_{\rm cdm}$ denotes a cold DM component of unspecified origin and does not include any contribution from thermally produced axions considered in this paper.}; the scalar amplitude of primordial perturbations $A_s$; the scalar spectral index $n_s$; the optical depth to reionization $\tau_{\rm reio}$; and the angular size of the horizon at recombination $\theta_s$. 
To account for the effects of axions on cosmology we consider two additional parameters: their mass $m_a$ and the inverse strength of their interaction $f/|C|$, which governs the shape of the axion distribution function $f_a(x,q)$.
For neutrinos, we assume three degenerate massive species with a total mass $\sum m_\nu \geq 0.06$ eV, consistent with the assumptions made by the Planck Collaboration~\cite{Planck:2018vyg}. Altogether, our extended model $\Lambda$CDM+$m_a$+$f/|C|$+$\sum m_\nu$ includes the six baseline $\Lambda$CDM parameters, the sum of neutrino masses (constrained from below as $\sum m_\nu \gtrsim 0.06$\,eV \cite{DeSalas:2018rby}), and two axion parameters. This amounts to nine free parameters in total, or eight when the axion mass is fixed. 
For our numerical analysis we use the Boltzmann-Einstein solver \texttt{CLASS}\footnote{\url{https://github.com/lesgourg/class_public}}~\cite{CLASS2, CLASS4} to calculate the cosmological observables. To include the non-linear corrections to power spectra we use the \texttt{halofit} routine~\cite{Takahashi:2012em, Ali-Haimoud:2012fzp} incorporated into \CLASS. We implement the axions originating from lepton processes as an additional non-cold dark matter degree of freedom and modify the code to include axion momentum distribution obtained by solving the fBE \cite{Badziak:2024qjg} (see \ref{app: fits} for details). Note that we investigate one axion coupling at a time, setting the others to zero to obtain conservative bounds for each channel.

To constrain the parameter space of our model, we perform Markov Chain Monte Carlo (MCMC) analyses by connecting \texttt{CLASS} to the \texttt{MontePython} sampler\footnote{\url{https://github.com/brinckmann/montepython_public}}~\cite{Audren:2012wb, Brinckmann:2018cvx}. We consider two complementary strategies: 1) fixing the axion mass $m_a$ and varying the other parameters (see Section~\ref{sec:results}); 2) varying all the parameters with $m_a$ having a log-flat prior (see \ref{app: full-scans}). We choose to fix the mass of axions for several reasons: first of all, when the mass is varied across a few orders of magnitude the MCMC scan tends to concentrate in the regions where the likelihood is the highest (below 0.1 eV), which complicates the determination of the $f/|C|$ bound at higher masses. Furthermore, the constraint in this region is already well approximated by the $\dNeff$ limit. Fixing the mass allows us to probe the whole mass region more robustly. Second, from the particle physics model-building perspective the axion mass is often a \textit{prediction} rather than a free parameter (unlike the coupling), hence the first strategy is more aligned with setting cosmological constraints on particular axion models. 

The sum of neutrino masses consists of the contributions from three degenerate massive species, each with a mass $m_\nu \in [0.02,\,0.1]$ eV, such that $\sum m_\nu \geq 0.06$ eV. The axion interaction parameter $f/|C|$ is scanned with a log-flat prior, taking $\log_{10}(f/|C|/{\rm GeV})$ in the range $[4.0,\,8.0]$ for LFC $\tau$ couplings, $[6.0,\,8.0]$ for LFC $e$ 
and $\mu$ couplings, $[7.0,\,9.0]$ for LFV $\mu$ couplings, and $[4.7,\,9.0]$ for LFV $\tau$ couplings. Note that for large values of $f/|C|$ axion abundance becomes negligible, see Fig. \ref{fig:dNeff_all}. Consequently, the data loses sensitivity to this parameter, leading to flat posteriors and prior-dependent limits. 
In \ref{App: Prior Dependence}, we use the $\mu$ LFC and $\tau$ LFV couplings as examples to illustrate how different choices for the upper prior limit impact our constraints, demosntrating that the choice reported in this section leads to conservative limits. Planck nuisance parameters are varied with the standard Gaussian priors provided in the official likelihoods. All remaining baseline cosmological parameters are assigned flat priors sufficiently broad that the resulting posteriors are entirely likelihood-dominated.

Our constraints are derived from a combination of CMB and
BAO data. For the CMB we use the 2018 Planck likelihoods, including high-$\ell$ TT, TE, and EE spectra, the low-$\ell$ temperature and polarization data, as well as the CMB lensing likelihood~\cite{Planck:2019nip, Planck:2018lbu}. For 
BAO we include the most recent results from the DESI DR2 release~\cite{DESI:2025zgx}. The MCMC chains are obtained via the Metropolis-Hastings algorithm and we consider them to be converged once they reach the Gelman-Rubin convergence criterion $R-1<0.02$. Posterior distributions and credible intervals are obtained using \texttt{GetDist}\footnote{\url{https://github.com/cmbant/getdist}}~\cite{Lewis:2019xzd}. 

\section{Results}
\label{sec:results}
Let us now discuss the results of our cosmological analysis in the form of constraints on 
axion-lepton couplings for lepton-flavor conserving (LFC) and lepton-flavor violating (LFV) processes separately. In addition, we analyse an approximate bound on warm DM component,
which can be relevant for axions with masses $m_a\gtrsim\mathcal{O}(10)$~eV.

\subsection{Lepton-flavor conserving (LFC) interactions}

\begin{figure}[h]
    \centering
    \includegraphics[width=\columnwidth]{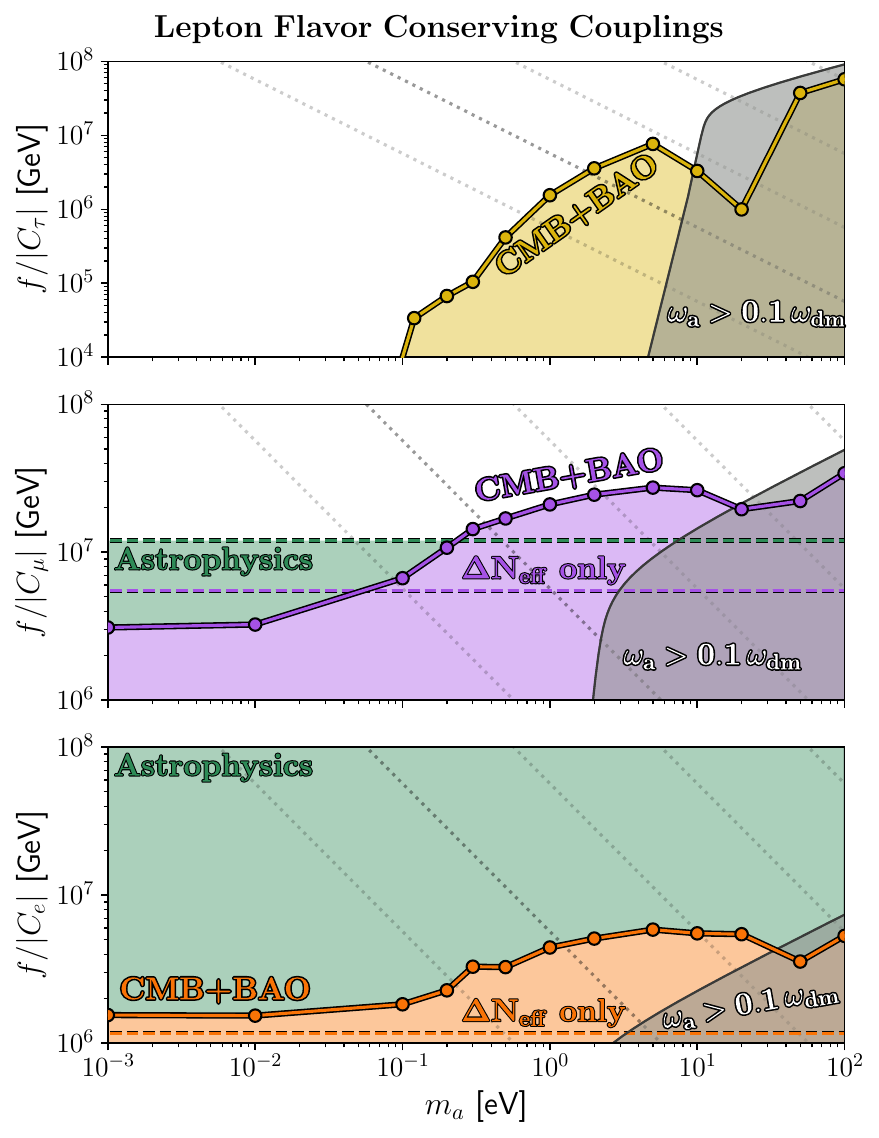}
    \caption{
    Summary of constraints on LFC axion couplings $f/|C_l|$ as a function of axion mass $m_a$. The colored points connected with solid lines indicate the 95\% C.L lower bounds on the couplings derived from CMB and BAO measurements (Planck 2018 + DESI data). The dashed lines of the same color represent the mass-independent bound on $f/|C_l|$ from the simplified $\Delta N_\mathrm{eff}$ analysis. The grey region is excluded by the approximate wDM constraint (see Sec.~\ref{Sec: wDM}). Green shading denotes mass-independent astrophysical limits. The dotted lines indicate the cross-sections of the parameter space for the QCD axions with the values of $\log_{10} |C_l|= [-3, -2, -1, 0, 1]$, from right to left
    ($|C_l|=1$ is highlighted by a darker color). See the text for more details. 
    }
    \label{fig:lfc_mass_limits_scatter_plot}
\end{figure}

In Figure~\ref{fig:lfc_mass_limits_scatter_plot} we present the 95\% C.L. limits on $f/|C_i|$ as a function of axion mass for all LFC couplings, shown as colored points connected by solid lines and labeled as CMB+BAO. Each point is the result of a scan with a fixed value of axion mass, from which we obtain the 95\% C.L. bound on $f/|C_i|$, and the line simply interpolates between these points. 
For small axion masses $m_a\lesssim \mathcal{O}(0.1)$ eV axions behave predominantly as dark radiation and are therefore mainly constrained by their contribution to $\Delta N_\mathrm{eff}$.
The cosmological bound increases with the axion mass in all channels up to a few eV. As the mass grows, so does the axion energy density. 
Thus, to avoid an overproduction of relativistic axions the interaction strength at higher masses must be weaker.  
Note that in the case of the $\tau$ coupling we do not obtain any meaningful 95\% C.L. constraints for masses $m_a<0.12$ eV as the axion abundance in this case is relatively small in comparison to other channels.

As the axion mass grows beyond $\mathcal{O}(10)$ eV, the cosmological constraints begin to weaken\footnote{The bounds resulting from our calculations -- with and without the \texttt{halofit} procedure for non-linear corrections to power spectra -- display a difference of up to a few tens of percent in the considered mass range. Nevertheless, we expect that the actual constraints in this mass range are dominated by the Ly-$\alpha$ forest effects, see Sec.~\ref{Sec: wDM}.}.
As already observed in Ref.~\cite{Caloni:2022uya}, in this mass regime, the axions behave predominantly as non-relativistic matter at recombination ($T\approx 0.3$ eV), so their contribution to $\Delta N_\mathrm{eff}$ is reduced and the bound on the energy density of axions is relaxed. Consequently, the limits on $f/|C_i|$ become less stringent in this region, as previously noted in Ref.~\cite{Caloni:2022uya} for axion couplings to photons and gluons. In particular this behaviour is most pronounced for the $\tau$ coupling. This is explained by the relatively weak dependence of the axion energy density on the value of $f/|C_\tau|$, as can be seen in Fig.~\ref{fig:dNeff_all}, which implies that even small relaxation of the bound on the axion energy density leads to a significantly weaker constraint on $f/|C_\tau|$ for $m_a \gtrsim 10$ eV.
Finally, in the limit of large axion mass values $m_a \gtrsim 50$ eV the bound grows stronger again as the abundance of massive axions approaches the observed relic density of CDM or even starts to exceed it. 

Around $m_a \sim 10$ eV our constraint becomes comparable to an approximate bound on the allowed fraction of warm DM coming from Ly-$\alpha$ forest observations. Note that we only extrapolate the results of hydrodynamical simulations for particles with $m \sim \mathcal{O}$(keV), so this bound should be treated simply as an indication that this region is potentially excluded (see Sec.~\ref{Sec: wDM} for more details).

For the sake of comparison we show the simplified cosmological bounds derived exclusively from the $\Delta N_\mathrm{eff}$ constraint with the dashed horizontal lines of the same color as the CMB+BAO constraint. These are computed under the assumption that axions remain fully relativistic at recombination and can be compared with the upper limit on $\Delta N_\mathrm{eff}\leq 0.3$ at 95\% C.L. from the Planck analysis 
(see Sec.~\ref{Sec:axion_production} and Fig.~\ref{fig:dNeff_all} in particular). Since the axion abundance depends solely on their interaction strength, these bounds are mass independent. The constraint is absent for the $\tau$ coupling in the plot as it would fall below the relevant parameter space $f/|C_\tau| < 10^4$ GeV.
While the limits obtained with each approach are very similar for small axion masses\footnote{The constraints on axion couplings stemming from $\Delta N_\mathrm{eff}\leq 0.3$ are slightly different than those derived by us from Planck and DESI data for $m_a\ll0.1$~eV.  
The inclusion of DESI results relaxes the upper bound on $\dNeff$, as was already observed e.g. in Ref.~\cite{Allali:2024cji}, while the choice of the prior (in combination with a steep dependence of $\dNeff$ on $f/|C|$ ) can result in a weaker bound comparing to the $\dNeff$ constraint.}, at higher masses our full analysis predictably provides stronger constraints than the simplified approach. This highlights the importance of properly accounting for the axion mass and for the transition from the relativistic to the non-relativistic regime.

In addition we show the complementary astrophysical constraints on the interaction strength as green dash-dotted lines and the shaded region of the corresponding color to highlight the excluded parameter space. The electron and muon couplings are constrained by the distinct astrophysical observations: the observed shape of the white dwarf luminosity function~\cite{MillerBertolami:2014rka} and the energy-loss argument for SN1987A~\cite{Caputo:2021rux}, respectively. These yield the following 95\% C.L. limits:
\begin{align}
    \frac{f}{|C_e|}\gtrsim 2\times 10^9 \text{ GeV}, && \frac{f}{|C_\mu|}\gtrsim 1.2\times 10^7 \text{ GeV}.
\end{align}
In contrast, there is no strong astrophysical bound for $\tau$ coupling as this type of leptons typically plays a very small role in astrophysical processes. The axion-$\tau$ coupling can be constrained indirectly by the astrophysical bound on axion-electron coupling since the latter is generated by the $\tau$ coupling at one-loop level. This results in a lower bound on $f/|C_{\tau}|$ at the level of a few times $10^4$~GeV~\cite{DEramo:2018vss}. We do not show this bound in Figure~\ref{fig:lfc_mass_limits_scatter_plot} since its exact value depends on the UV cut-off of low-energy theory, so it is model-dependent. 

Our cosmological analysis provides the most stringent limit to date on the $\tau$ coupling for axion masses above $\mathcal{O}(0.1)$~eV and the values of $f/|C_{\tau}|$ of up to $\mathcal{O}(10^7)$~GeV are excluded for $m_a$ in the eV range. For the $\mu$ coupling our results are competitive with the astrophysical constraint and for axion masses above $0.2$~eV our bound can even surpass it, excluding $f/|C_{\mu}|$ below a few times $10^7$~GeV. On the other hand, the astrophysical bound for the electron coupling is much stronger than that from cosmology.

Our results have also implications for the QCD axion. The dotted grey lines in Fig.~\ref{fig:lfc_mass_limits_scatter_plot} show the cross sections of the parameter space that correspond to the QCD axions for which the relation in Eq.~\eqref{eq: axion zero temperature mass} holds for various values of $C_l$. 
Thus, for QCD axions with $|C_l| \leq 1$ our results provide stronger constraints on $f$ compared to the mass-independent $\dNeff$ limits for all channels and the strongest constraints in general for axion couplings to muons and tau leptons. In particular, for $C_\tau=1$ cosmology excludes the values of $m_a\gtrsim2$~eV, or equivalently, $f\lesssim3\times10^6$~GeV, which would otherwise be unconstrained if naive $\dNeff$ constraint was applied. For $C_\mu=1$, $m_a\gtrsim0.4$~eV and $f\lesssim1.5\times10^7$~GeV are excluded, which is an improvement of about a factor of three with respect to the constraint from $\dNeff$ that neglects the mass of the axion. For $C_e=1$, the lower bound on $f$ is improved to about $5\times10^6$~GeV, but it is still weaker than the astrophysical bound by almost three orders of magnitude.

\subsection{Lepton-flavor violating (LFV) interactions}

\begin{figure}
    \centering
    \includegraphics[width=\columnwidth]{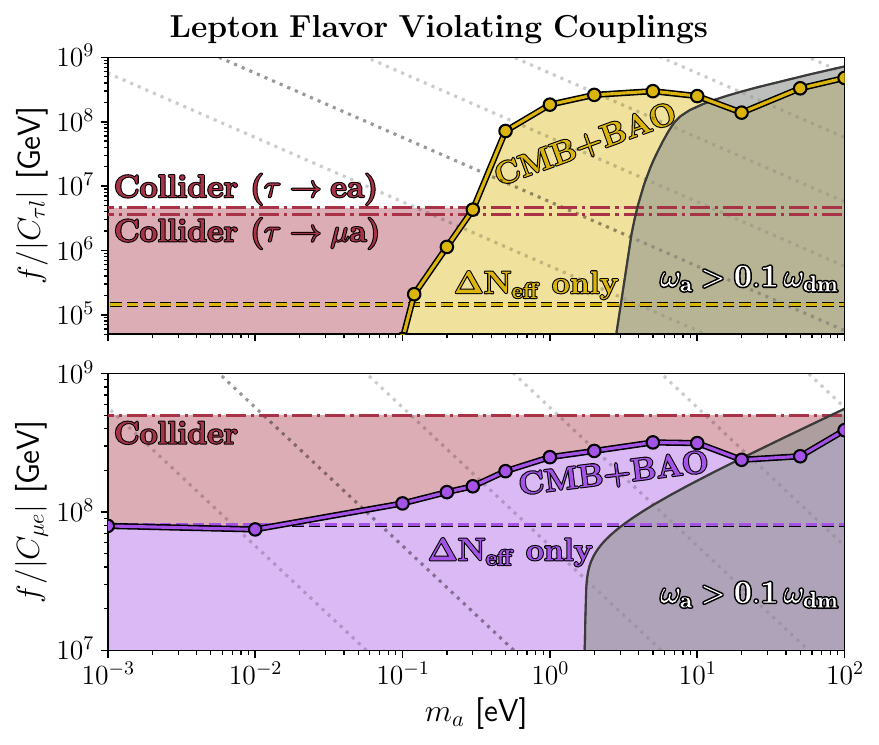}
    \caption{Same as Fig.~\ref{fig:lfc_mass_limits_scatter_plot}, but for LFV axion couplings.
    }
    \label{fig:lfv_mass_limits_scatter_plot}
\end{figure}

In Figure~\ref{fig:lfv_mass_limits_scatter_plot} we present the 95\% C.L. limits for the LFV couplings. As the axion distribution functions for $\tau$ decays into electrons and muons are virtually indistinguishable we display only one plot with the constraints on $f/|C_{\tau l}|$, where $l$ is $\mu$ or $e$. Our limits demonstrate qualitatively similar behaviour to the ones obtained for LFC couplings and can be explained via the same physical reasoning. 
We also show the $\Delta N_\mathrm{eff}$ limits, and the comparison between these simplified constraints and our full cosmological constraints reaches the same conclusion as for the LFC case: the limits at low masses are comparable, while at higher masses the simplified limits are much weaker. 
Instead of the astrophysical bounds as in the LFC case, the LFV couplings are constrained by the collider measurements and the respective limits are shown as red dash-dotted lines.\footnote{Astrophysical constraints on LFV axion couplings are weaker than the collider ones, see e.g. Refs.~\citep{Calibbi:2020jvd, Zhang:2023vva} and a recent review~\citep{MartinCamalich:2025srw} of constraints on FV axion couplings.} For the tau couplings, the limits are set by Belle-II~\citep{Belle-II:2022heu}; for the muon couplings, they are derived from the TRIUMF and TWIST experiments~\citep{PhysRevD.34.1967, TWIST:2014ymv}.
The corresponding current bounds at 95\% C.L. are:
\begin{equation}
\label{eq:lfv-couplings}
\begin{gathered}
\frac{f}{|C_{\tau e}|} \gtrsim 4.6\times 10^6~\mathrm{GeV},
\qquad
\frac{f}{|C_{\tau\mu}|} \gtrsim 3.6\times 10^6~\mathrm{GeV}, \\
\frac{f}{|C_{\mu e}|} \gtrsim 5\times 10^8~\mathrm{GeV}.
\end{gathered}
\end{equation}
The collider bound on $C_{\mu e}$ depends on the chirality structure. The bound quoted above assumes  purely left-handed coupling, corresponding to $C_{\mu e}^V=-C_{\mu e}^A$, which is weaker than that for purely right-handed coupling by a factor of about 2.5~\cite{Calibbi:2020jvd}.   

The most significant impact of the axion mass on the cosmological bound can be seen for the LFV axion couplings involving $\tau$ lepton.
Our cosmological bounds for $\tau$ couplings are competitive with the collider limits and even supersede them for axion masses above 0.3~eV. For masses above about 1~eV the lower bound on $f/|C_{\tau l}|$ exceeds $10^8$~GeV, which is an order of magnitude stronger than a projected future bound from Belle-II with 50ab$^{-1}$ of data estimated to be $f/|C_{\tau l}|\gtrsim2\times10^7$~GeV~\cite{Badziak:2024szg}. The cosmological bound on the QCD axions with $|C| \leq 1$ is the strongest of all for $\tau$ couplings, which for $C_{\tau l}=1$ corresponds to $m_a\lesssim0.4$~eV, or equivalently, $f\gtrsim2\times10^7$~GeV. On the other hand, for LFV axion-$\mu$-$e$ coupling the collider bounds remain stronger than cosmological constraints over the entire mass range considered in our analysis with \texttt{CLASS}. Nevertheless, for $m_a\gtrsim100$~eV we expect that the constraints on axions from Ly-$\alpha$ forest may become stronger than the collider ones.  

Let us also compare our results with those of Ref.~\cite{Xu:2021rwg}, which investigated cosmological constraints on light, but massive thermal relics that decoupled from the SM plasma before the EW phase transition. Even though Ref.~\cite{Xu:2021rwg} adopted instantaneous decoupling approximation it is still possible to approximately translate their upper bounds on the temperature of light relics into bounds on axion couplings. We find a good agreement for axion masses up to 1~eV. For larger axion masses, up to 10~eV, our bounds are somewhat weaker than those inferred from the results of Ref.~\cite{Xu:2021rwg}. While a strict comparison of the results is not possible, as we used the latest DESI data instead of the BOSS data employed in Ref.~\cite{Xu:2021rwg}, they argue that the inclusion of additional weak-lensing data~\cite{Heymans:2013fya} and, to a lesser extent, of the full-shape galaxy data improves the upper bounds on the energy density of light relics by a factor of a few. Thus, our bounds for axion masses above 1~eV should be regarded as conservative. We expect that incorporating the weak-lensing data may improve bounds on some axion couplings by up to a factor of a few in this mass range. We leave a detailed investigation of this effect for future work.        

\subsection{Warm DM constraint}
\label{Sec: wDM}

\begin{figure*}
    \centering
    \includegraphics{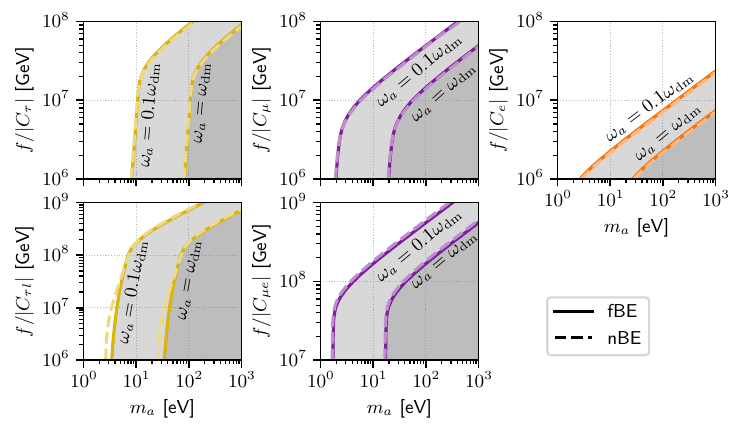}

    \caption{
    Constraints on the axion decay constant $f/|C|$ as a function of the axion mass $m_a$ from the axion contribution to the present-day DM abundance, $\omega_{\rm dm}=0.12$. Each panel corresponds to a distinct axion-lepton interaction with LFC (LFV) interactions shown on the upper (lower) panels.   
    Solid lines correspond to the calculation of the axion abundance via fBE approach, while the dashed lines correspond to the nBE approach.
    We show two contours of the present-day axion energy density: \( \omega_a = \omega_{\rm dm} \) (CDM bound) and \( \omega_a = 0.1\omega_{\rm dm} \) (wDM bound).  The shaded region below the $\omega_a = 0.1\omega_{\rm dm}$ indicates the parameter space that we estimate to be excluded due to overproduction.
    }
    \label{fig:warm-DM-constraints}
\end{figure*}

In addition to the impact on the early structure formation, that can be probed by CMB and BAO measurements, axions with masses $m_a \lesssim \mathcal{O}(10)$~keV are also expected to have an effect on the late-time structure formation probed by Ly-$\alpha$ forest observations. In other words, it means that warm DM limits can be applied to axions and are expected to constrain a region of the parameter space that we probe in our analysis more robustly. While the former effects on the cosmological observables can be reliably computed in the linear regime using \CLASS, the latter effects require hydrodynamical simulations. We use the results of these simulations from the literature in order to estimate the wDM constraints on the axion abundance. These constraints can be conveniently expressed in terms of the fraction of 
axions in the total dark matter energy density $f_{\rm wdm} \equiv \omega_{a}/\omega_{\rm dm}$, where $\omega_a \equiv \Omega_a h^2$ is the axion contribution to the energy density of the Universe.  

The results of Ref.~\cite{Garcia-Gallego:2025kiw} suggest that below $m_a \sim 5$~keV the upper limit on $f_{\rm wdm}$ decreases and reaches $f_{\rm wdm} = 16\%$ at 1~keV, which is the minimum of the mass range in that analysis. The authors state that the upper limit should keep decreasing towards smaller masses, however it is expected to ultimately reach a plateau due to the fact that the uncertainties of the current Ly-$\alpha$ data limit the ability to constrain light thermal relics more tightly. Ref.~\cite{Baur:2017stq} reaches a qualitatively similar conclusion and states that any mass of a new particle is allowed by the Ly-$\alpha$ data as long as it constitutes less than $10\%$ of the total DM density. Thus, we conservatively estimate the wDM constraint on the axions with the masses $m_a < 1$ keV as $f_{\rm wdm} \leq 0.1$. 

For the range of axion masses in our analysis one can confidently use the non-relativistic approximation for the present-day axion energy density

 \begin{equation}
    \rho_a(T_0) \simeq m_a \cdot n_a(T_0) \, ,
    \label{eq:rho_a_nonrel}
\end{equation}
where $n_a$ is the axion number density and $T_0$ is the present temperature of the Universe (CMB). In the fBE approach that we concentrate on in this paper the present-day axion number density can be calculated from the axion distribution function $f_a$ as follows 

 \begin{equation}
    n_a(T_0) = 
    \frac{g_a}{2\pi^2} T_0^3 \left(\frac{h_{s}(T_f)}{h_{s}(T_{0})}\right)^{-1} \int_0^{\infty} \tilde{q}^2 f_a(\tilde{q}) \, d\tilde{q},
\end{equation}
where $\tilde{q} \equiv q \cdot (h_s(T_0) / h_s(T_f))^{-1/3}$ is the reduced momentum coordinate which preserves the same shape of the axion distribution function across the entropy density evolution in the expanding Universe in the absence of axion interactions \cite{Badziak:2024qjg} and \( T_f \) is the temperature at which $f_a(q)$ is obtained\footnote{In our analysis we trace the evolution of $f_a(x,q)$ until $x = 20$, at which the axion comoving abundance is firmly established.}.
In the nBE approach the present-day axion density is obtained straightforwardly from the solution of the Boltzmann equation for the comoving abundance of axions $Y_a$ as 
\be
n_a(T_0) = Y_a \cdot s_0 \, ,
\ee
where $s_0$ is the present-day CMB entropy density. 

With these expressions we trace which combinations of $m_a$ and $f/|C|$ (that solely determines the axion abundance in the thermal scenario) result in \( \omega_a = 0.12 \) and \(\omega_a = 0.012 \). The first value corresponds to the whole observed DM abundance, while the second value corresponds to $10\%$ of the observed DM abundance, thus comprising the strong bound on the axion abundance from overclosing the Universe and a conservative estimate of the bound from Ly-alpha constraints. These bounds are presented in Fig.~\ref{fig:warm-DM-constraints} for different axion-lepton production channels. Each bound is computed for both fBE and nBE approaches.
The bounds in the plots constrain the regions of the axion parameter space located below the contours. 
The lower $m_a$ is, the higher the allowed comoving density of axions (see Eq.~\eqref{eq:rho_a_nonrel}). Consequently, the corresponding value of $f/|C|$  required to reproduce this density is reduced, which explains the downward trend of the constraint curves shown in the plots. 
The knees of the curves indicate the transition from freeze-in regime at large $f/|C|$ to freeze-out regime at small $f/|C|$. The only substantial difference between the nBE and fBE constraints is observed for $\tau$ decays, which is due to a smaller axion number density found with the fBE approach. This is basically the consequence of the decay kinematics (large non-thermal corrections due to the mass difference) and the rapid change of entropy degrees of freedom around the freeze-out\footnote{For more details see Ref.~\cite{Badziak:2024qjg}, especially Fig.~2 and the text around it.}.

A recent work~\cite{DEramo:2025jsb} investigated in detail the constraints on the mass of axions playing the role of warm DM, while taking into account the phase-space distributions of axions produced in leptonic channels via freeze-in. The lower bound on the axion mass was found to be between 20 and 40~keV with the exact value dependent on the axion production channel. Ref.~\cite{DEramo:2025jsb} also concluded that axions produced thermally in leptonic channels can play the role of DM\footnote{In an earlier work~\cite{Panci:2022wlc}, a detailed phenomenological analysis of thermally produced axion DM via the decay $\mu \to e a$ was carried out.    } except for the case when axion couples to electrons, in which axion DM is excluded by X-ray constraints because of too large an axion-photon coupling induced at one loop by the axion-electron coupling. Figure~\ref{fig:warm-DM-constraints} shows complementary results for $m_a<1$~keV for which thermally produced axions cannot explain all the observed dark matter, but are stable enough to avoid X-ray constraints and the strongest bounds are those from Ly-$\alpha$ forest constraints, which we approximate in this mass region demanding $\omega_a<0.1\omega_{\rm dm}$. We see that the lower bounds on $f/|C|$ for LFC axion couplings to muons and taus exceed $10^8$~GeV, while those for LFV axion couplings exceed $10^9$~GeV for axion masses approaching 1 keV and are stronger than astrophysical and collider constraints. The lower bound on $f/|C_e|$ for $m_a\approx1$~keV exceeds $10^7$~GeV, but it is still much weaker than the astrophysical bound.

\section{Conclusions}
\label{sec:conclusions}

We have investigated cosmological constraints on axion-lepton interactions using state-of-the-art computations of axion momentum-distribution functions arising from axions thermally produced in the SM bath.  We used the CMB measurements from Planck together with the BAO results from DESI DR2 to constrain both LFC and LFV axion couplings. Our results strengthen the existing cosmological bounds, that have been derived using upper bounds on $\dNeff$, in the range of axion masses between about 0.1~eV and 1~keV. 

\begin{figure}
    \centering
    \includegraphics{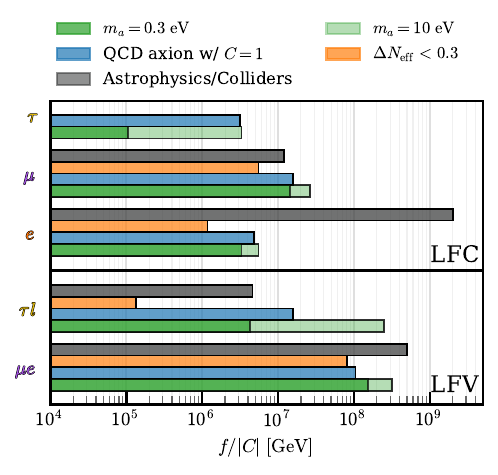}
    \caption{Summary of 95\% C.L. bounds on $f/|C|$ for LFC and LFV axion-lepton couplings. The orange bars represent the weakest cosmological limits based on the upper bound on $\Delta N_{\mathrm{eff}} < 0.3\,$ from Planck. Blue bars show the bounds for the axion decay constant for a QCD axion with $C=1$, while green bars illustrate two representative scenarios with non-vanishing axion mass, $m_a=10$~eV and $m_a=0.3$~eV. The strongest non-cosmological constraints are represented by the dark–grey bars and correspond to astrophysical (collider) bounds for LFC (LFV) couplings. }
    \label{fig:axion_limits}
\end{figure}

Figure~\ref{fig:axion_limits} exemplifies the impact of non-vanishing axion mass on the constraints on axion-lepton couplings. We compare the constraints derived from the Planck upper bound on $\dNeff<0.3$, which is valid under the assumption of negligible axion mass, with the constraints for $m_a=0.3$~eV and $m_a=10$~eV. 
We see that already for $m_a=0.3$~eV, when the axion mass is close to the recombination temperature, our constraints on axion-lepton couplings are significantly stronger than those derived from the upper bound on $\dNeff$. For $m_a=10$~eV, which is in the ballpark of axion masses maximizing the constraints, 
the bounds are improved by a factor of about 5 for LFC axion couplings to electrons and muons and LFV ones to muon and electron, while the bound on LFC and LFV axion couplings involving tau leptons are improved by many orders of magnitude. 
For axion masses above about 10~eV we expect that the 
cosmological constraints are enhanced by the Ly-$\alpha$ forest observations. 
The derivation of precise constraints from it is beyond the scope of this paper, but we derive approximate constraints relying on the computation of axion abundance. These constraints become stronger with increasing $m_a$ and become the strongest for $m_a\sim\mathcal{O}(1)$~keV.

For large enough masses, the cosmological constraints are also typically stronger than those from colliders and astrophysics. Cosmology provides the most stringent constraints on LFC axion couplings to taus and muons for axion masses above about 0.1 and 0.3~eV, respectively. The lower bounds on $f/|C_\tau|$ and $f/|C_\mu|$ range from about $10^4$ and $10^7$~GeV, respectively, for $m_a\sim\mathcal{O}(0.1)$~eV up to about $10^8$~GeV for $m_a\sim\mathcal{O}(1)$~keV, where the additional strong constraints on axion are expected from Ly-$\alpha$ forest.
The constraints on axion-electron couplings are not competitive with astrophysical constraints and the lower bound from cosmology on $f/|C_e|$ varies from about $10^6$~GeV for very light axions up to about $10^7$~GeV for $m_a\sim\mathcal{O}(1)$~keV.

For LFV axion couplings we found the lower cosmological bound on $f/|C_{\tau e}|$ and $f/|C_{\tau \mu}|$ to be stronger than the collider limits from Belle-II for axion masses above about 0.3~eV and this bound exceeds $10^9$~GeV for $m_a\sim\mathcal{O}(1)$~keV. Since the collider bounds on LFV axion couplings to muon and electron are very strong, cosmology cannot compete with them for light axions, but for axion masses above about 100~eV cosmological constraints become stronger and the lower bound on $f/|C_{\mu e}|$ exceeds $10^9$~GeV for $m_a\sim\mathcal{O}(1)$~keV.

In our analysis we chose priors on the axion couplings that lead to conservative lower bounds. We should, note, however that that these bounds become stronger if the prior range for $f/|C_i|$ extends to larger values, see Figure~\ref{fig:prior_comparison}. 

Our results can be also applied to the QCD axion. We see in Figure~\ref{fig:axion_limits} that for axion-lepton couplings with $C\sim\mathcal{O}(1)$ the bounds on the axion decay constant are stronger than the simplified ones derived from the upper bound on $\dNeff$. The strongest cosmological constraints are for the QCD axions with sizable LFC couplings to muons or LFV couplings involving tau, for which the values of the axion decay constant above $10^7$~GeV are excluded. Although astrophysical constraints on the QCD axion couplings to nucleons often lead to stronger bounds on $f$, in some astrophobic QCD axion models~\cite{DiLuzio:2017ogq,Badziak:2024szg} our cosmological constraints provide the strongest lower bound on the axion decay constant. 

\section*{Acknowledgments} 

The authors would like to thank Daniele Perri and Olga Garcia-Gallego for useful discussions. This work was partially supported by the National Science Centre, Poland, under research grant no. 2020/38/E/ST2/00243.

\appendix

\section{Implementation of axion distribution functions in \texttt{CLASS}} \label{app: fits}
\begin{figure}[t]
    \centering
    \begin{subfigure}{\columnwidth}
        \centering
        \includegraphics{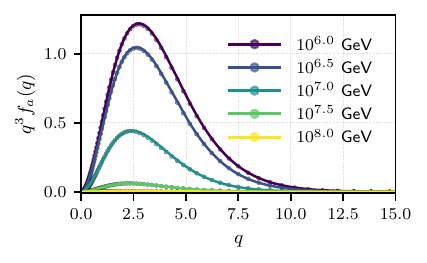}
    \end{subfigure}
    \begin{subfigure}{\columnwidth}
        \centering
        \includegraphics{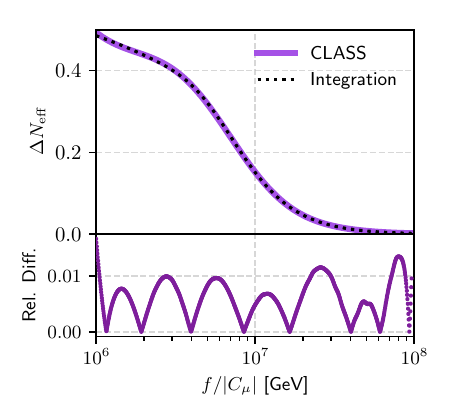}
    \end{subfigure}
    
    \caption{\textbf{Upper Panel:} Comparison of the axion momentum distributions resulting from LFC $\mu$ interactions for different values of $f/|C_\mu|$ and their fitting functions. The points correspond to the distributions obtained numerically and the solid lines to the fit, as explained in the text. \textbf{Lower Panel:} $\dNeff$ calculated by integrating the numerical distributions (black dotted line) and obtained from \CLASS~with the procedure explained in this appendix (purple line). The lower panel shows the relative differences between these two approaches. 
    }
    \label{fig:fBE fits}
\end{figure}

In this Appendix we describe in detail how we implement axion distribution functions in \texttt{CLASS}. The fBE for the axion energy distributions $f_a(q)$ is solved numerically for a range of discrete values of axion interaction strength $f/|C|$ as in Ref.~\cite{Badziak:2024qjg}, thus we obtain a set of 
tabulated distributions for each value considered.
While \texttt{CLASS} can in principle process such tabulated results
for fixed values of $f/|C|$,
this is insufficient for our purposes: to constrain the axion coupling with cosmological data we need to explore arbitrary values of $f/|C|$ in our MCMC analyses. 
To achieve this we perform analytical fits to the numerically obtained distributions. We find that they are well approximated by the following functional form
\begin{equation}
    q^2f_a(q)=\sqrt{1+q^2} \; q^d\left(\exp{(A\sqrt{1+q^2}-M)}+k\right)^{-1} \, ,
    \label{eq: analytical distribution}
\end{equation}
where $q\equiv E_a/T$ and the parameters $\alpha \equiv \{ A$, $M$, $k$, $d \}$ are smooth functions of the interaction strength. In particular, they can be expressed as polynomials in $\log{f/|C|}$: 
\begin{equation}
    \alpha_i(f/|C|)=b_0+b_1\log{f/|C|}+b_2(\log{f/|C|})^2+...
    \label{eq: distribution parameters fit}
\end{equation}
with $b_i$ being some coefficients obtained via fitting. 
Therefore, our procedure consists of two steps:
\begin{enumerate}
    \item For each value of $f/|C|$, we fit the function \eqref{eq: analytical distribution} to the numerical distribution, obtaining a set of parameters $\{ A, M, k, d \}$.
    \item The resulting parameters are then fitted using \eqref{eq: distribution parameters fit}, providing a continuous mapping between $f/|C|$ and the distribution function.
\end{enumerate}
As an illustration, in the upper panel of Fig.~\ref{fig:fBE fits} we present the results for the LFC axion-muon coupling. The points correspond to the numerical solution of the fBE and the solid lines to the fitted distributions. We see that the fitted functions agree very well with the numerical results. As a cross-check, the lower panel shows the resulting values of $\dNeff$: the dotted black line corresponds to the values obtained by integrating the numerical distributions directly, while the solid light purple line shows the results from \texttt{CLASS} obtained with our fitted functions. The lower panel shows the relative difference between the two approaches. Once again, we observe a good consistency between the results of using actual distributions and the results of using fitting functions, confirming that our implementation is suitable for MCMC scans.

\section{Results of full scans with varying axion mass} \label{app: full-scans}

\begin{figure*}[h!t]
    \centering
    \includegraphics[width=0.7\textwidth]{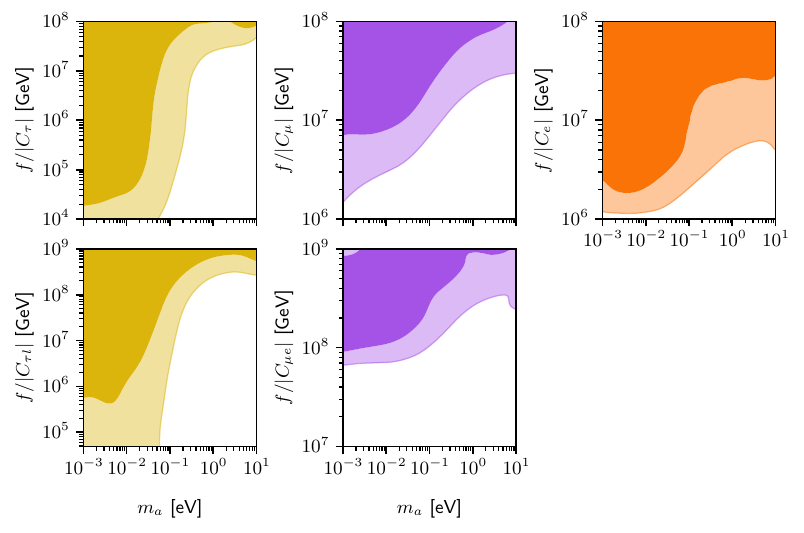}
    \caption{Posteriors of the full cosmological scans in the plane $f/|C|$-$m_a$ with varying axion mass $m_a$ for all considered channels of axion production. Darker and lighter color shadings indicate $1\sigma$ and $2\sigma$ regions respectively. 
    }
    \label{fig:full_scans_fa_ma}
\end{figure*}

In this Appendix we present the results of the full cosmological scans in which we treat $m_a$ as a free parameter and scan over it in the range $[10^{-3}, 10]$ eV. Our upper bound on the axion mass prior is only 10~eV, because for larger masses the constraints from Ly-$\alpha$ forest are expected to be stronger than those from CMB and BAO. Similarly to the analysis described in the main text, we consider our scans converged once they reach the Gelman-Rubin factor $R-1<0.02$. In Fig.~\ref{fig:full_scans_fa_ma} we show the preferred regions in the $f/|C|$-$m_a$ plane for LFC and LFV couplings obtained after marginalizing over all other cosmological and nuisance parameters. In Table \ref{tab:best-fit full} we list the best-fit values and $95\%$ limits for the scanned cosmological parameters. 

\begin{figure*}[h!t]
    \centering
    \includegraphics{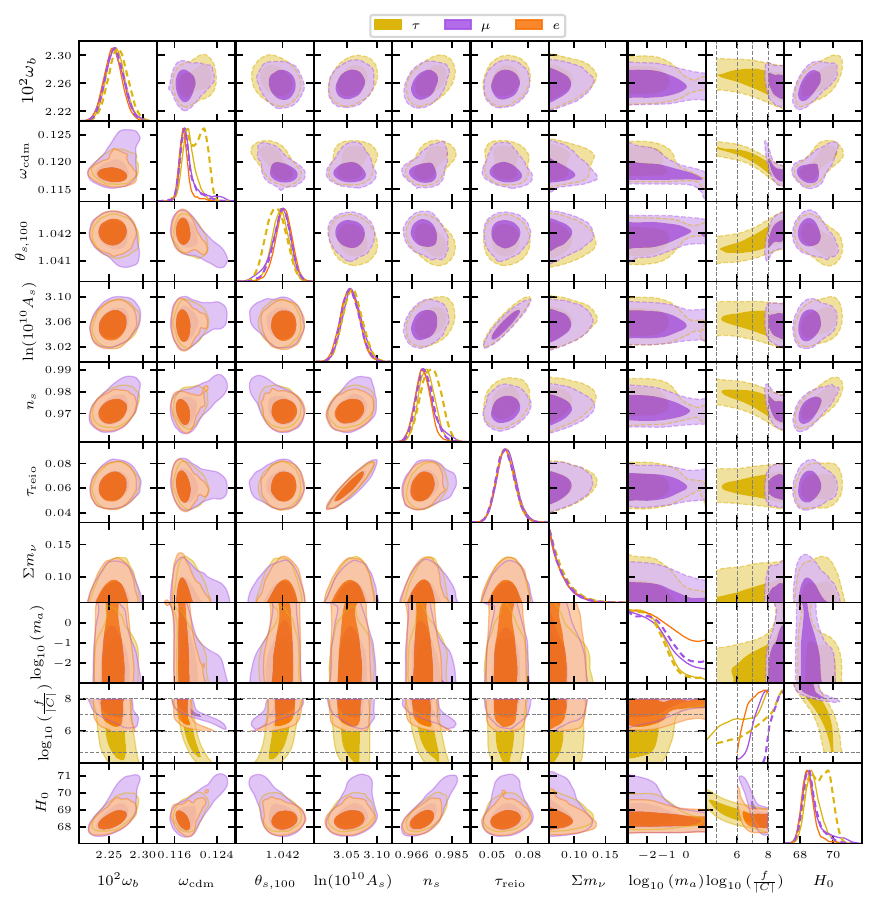}
    \caption{1D and 2D posterior plots for all the cosmological parameters and axion parameters considered in this work and corresponding to the scans with $m_a$ being a free parameter. The lower left half of the figure corresponds to the LFC couplings (solid lines on 1D plots), while the upper right half -- to the LFV ones (dashed lines on 1D plots). Darker and lighter color shadings indicate $1\sigma$ and $2\sigma$ regions respectively. The dashed lines in the plots involving $\log_{10}(f/|C|)$ denote the boundaries on the corresponding prior used in the scans. The units of $m_a$ and $\Sigma m_\nu$ are eV, the unit of $f$ is GeV, while $H_0$ is written in units of (km/s$\cdot$Mpc). }
    \label{fig:full_scans_all}
\end{figure*}

In Fig.~\ref{fig:full_scans_all} we show 1D and 2D posterior distributions (triangle plots) for all the cosmological parameters and axion parameters resulting from our full scans with varying axion mass. The lower left triangle corresponds to LFC couplings, while the upper right triangle corresponds to LFV couplings, and different colors denote different lepton modes. Darker and lighter shadings in 2D plots correspond to $1\sigma$ and $2\sigma$ regions respectively. In 1D plots the likelihoods for LFC couplings are plotted with solid lines and with dashed lines for LFV couplings. The dashed lines in 1D and 2D plots for $f/|C|$ parameter indicate the boundaries on the priors for different lepton channels. Note that $f/|C|$ vs. $m_a$ plots in this section should not be compared with our constraints described in the main text (e.g. Fig.~\ref{fig:lfc_mass_limits_scatter_plot}), as those results correspond to a different type of scans with fixed axion masses. 

\begin{table*}[h!t]
    \centering
    \arrayrulecolor{black}
    \resizebox{0.95\textwidth}{!}{
    \begin{tabular}{l >{\columncolor{coltau!40}}c >{\columncolor{colmu!40}}c >{\columncolor{coltau!40}}c >{\columncolor{colmu!40}}c >{\columncolor{cole!40}}c}
        \toprule
        \specialrule{.05em}{0em}{0em}
        \textbf{Parameter} & 
        \textbf{$\tau$ LFV} & 
        \textbf{$\mu$ LFV} & 
        \textbf{$\tau$ LFC} & 
        \textbf{$\mu$ LFC} & 
        \textbf{$e$ LFC} \\
        \specialrule{.01em}{0em}{0em}
        \midrule
        \specialrule{.01em}{0em}{0em}
        {\boldmath$10^{2}\omega{}_\mathrm{b }$} & $2.264^{+0.028}_{-0.028}   $ & $2.259^{+0.027}_{-0.026}   $ & $2.260^{+0.027}_{-0.026}   $ & $2.258^{+0.030}_{-0.027}   $ & $2.256^{+0.027}_{-0.025}   $\\
        
        {\boldmath$\omega{}_\mathrm{cdm }$} & $0.1198^{+0.0032}_{-0.0032}$ & $0.1184^{+0.0038}_{-0.0025}$ & $0.1188^{+0.0028}_{-0.0022}$ & $0.1188^{+0.0058}_{-0.0028}$ & $0.1179^{+0.0026}_{-0.0021}$\\
        
        {\boldmath$\theta{}_{s, 100 }$} & $1.04176^{+0.00070}_{-0.00069}$ & $1.04196^{+0.00065}_{-0.00068}$ & $1.04191^{+0.00058}_{-0.00062}$ & $1.04191^{+0.00071}_{-0.00082}$ & $1.04203^{+0.00057}_{-0.00065}$\\
        
        {\boldmath$\ln 10^{10}A_{s }$} & $3.060^{+0.032}_{-0.030}   $ & $3.056^{+0.031}_{-0.028}   $ & $3.057^{+0.031}_{-0.028}   $ & $3.059^{+0.033}_{-0.030}   $ & $3.055^{+0.031}_{-0.029}   $\\
        
        {\boldmath$n_{s }         $} & $0.9750^{+0.0089}_{-0.0090}$ & $0.9723^{+0.0084}_{-0.0081}$ & $0.9731^{+0.0078}_{-0.0074}$ & $0.972^{+0.011}_{-0.0090}  $ & $0.9713^{+0.0074}_{-0.0075}$\\
        
        {\boldmath$\tau{}_\mathrm{reio }$} & $0.061^{+0.016}_{-0.014}   $ & $0.061^{+0.016}_{-0.014}   $ & $0.061^{+0.015}_{-0.014}   $ & $0.062^{+0.016}_{-0.014}   $ & $0.062^{+0.016}_{-0.014}   $\\[2pt]
        \specialrule{.01em}{0em}{0em}
        \midrule
        \specialrule{.01em}{0em}{0em}
        {\boldmath$\log_{10}{(m_a/\mathrm{eV})} $} & $< -0.263                  $ & ---                          & $< -0.225                  $ & ---                          & ---                         \\
        
        {\boldmath$\log_{10}{(f/|C|/\mathrm{GeV})}$} & ---                          & $> 7.98                    $ & ---                          & $> 6.58                    $ & $> 6.25                    $\\
        
        {\boldmath$\Sigma m_\nu \mathrm{\, [eV]}$} & $< 0.119                   $ & $< 0.114                   $ & $< 0.115                   $ & $< 0.111                   $ & $< 0.115                   $\\
        \specialrule{.01em}{0em}{0em}
        \bottomrule
    \end{tabular}
    }
    \caption{Best fit values and 95\% Credible Limits for the scanned cosmological parameters.}
    \label{tab:best-fit full}
\end{table*}

\section{Prior dependence of the results}
\label{App: Prior Dependence}
In this Appendix, we illustrate how the choice of prior upper bounds affects our limits on $f/|C|$. As Fig. \ref{fig:dNeff_all} demonstrates, large $f/|C|$ values correspond to low axion energy densities; therefore, we expect that the data becomes insensitive in this regime, causing this parameter's posterior to flatten. 

\begin{figure}[h!]
    \centering
    \begin{subfigure}{0.48\textwidth}
        \centering
        \includegraphics[width=\linewidth]{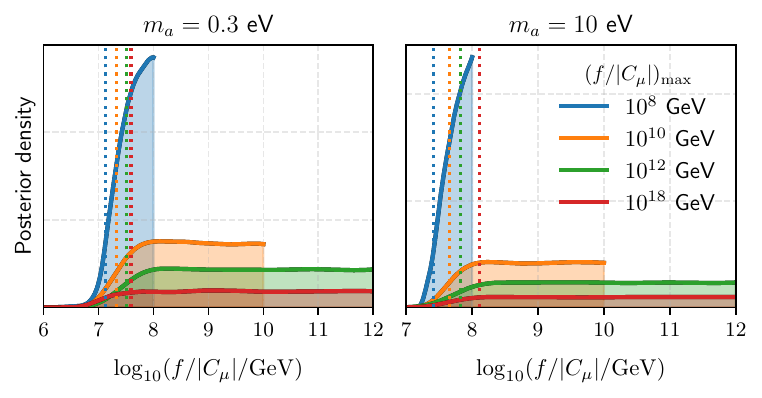}
        \caption{$\mu$ Flavor Conserving Coupling}
        \label{fig:priors_mu}
    \end{subfigure}
    \vfill
    \begin{subfigure}{0.48\textwidth}
        \centering
        \includegraphics[width=\linewidth]{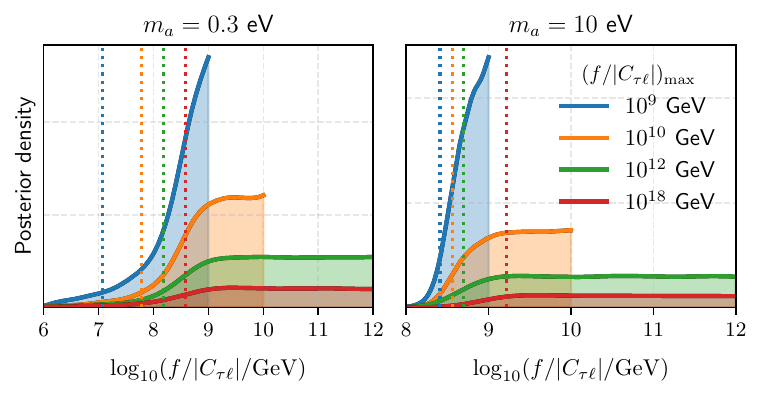}
        \caption{$\tau$ Flavor Violating Couplings}
        \label{fig:priors_tau}
    \end{subfigure}
    \caption{Posterior densities of $\log_{10}(f/|C|/\mathrm{GeV})$ for $\mu$ LFC and $\tau$ LFV couplings for axion masses $m_a=0.3$ and $10$ eV. Each distribution corresponds to a different choice for the upper prior bound. The vertical dotted lines denote the associated 95\% credible limits. }
    \label{fig:prior_comparison}
\end{figure}

To test this, we repeat our analysis using four different upper prior limits on $f/|C|$. Alongside the baseline limits used in the main text i.e. $10^8$ GeV or $10^{9}$ GeV depending on the channel, we extend our prior bounds to $10^{10}$ GeV, $10^{12}$ GeV and $10^{18}$ GeV. Figure \ref{fig:prior_comparison} shows the behavior of the posteriors for two representative axion masses ($m_a=0.3$ eV and $10$ eV) for $\mu$ FC and $\tau$ FV channels as an example.
For visual clarity, we cut the horizontal axis at $10^{12}$ GeV, although the red curve corresponding to upper prior bound at $f/|C|=10^{18}$ GeV extends further. As expected, the posteriors indeed flatten beyond $10^{8}$ GeV and $10^{9}$ GeV for $\mu$ and $\tau$ channels, respectively. We also plot the 95\% credible limits as vertical dotted lines for each prior choice. We see that the limits shift towards larger $f/|C|$ values, since the extended prior volume results in a greater share of the probability mass in the flat region of the parameter space. Note that the impact of the prior choice is more pronounced for $\tau$ LFV couplings. As axion production via this channel is less efficient and has milder dependence on $f/|C|$, larger coupling values can remain unconstrained by the data. However, increasing the upper prior limit introduces a large volume of the parameter space corresponding to a negligible axion abundance, which shifts the posterior distribution more significantly than in other channels, for which low values of $f/|C|$ are constrained anyway. 

Consequently, we restrict the prior bounds in our analysis to $10^{8}$ GeV and $10^{9}$ GeV, depending on the channel. This ensures that our analysis remains focused on the parameter space that is the most sensitive to the data, allowing us to report conservative limits. 

\section{CMB power spectra for $\tau$ decays}
\label{app:power_spectra}

In this Appendix we show the predictions from various axion models that we include in our MCMC scans on the example of axion-tau LFV processes. We choose this particular channel for the purpose of demonstration as it usually yields the most pronounced effects comparing to other axion-lepton channels. 

We consider two observable quantities: temperature power spectrum $C^{TT}_\ell$ and lensing power spectrum $C^{\phi\phi}_\ell$. 
As the main motivation of our analysis is to set constraints on axion models, it is more convenient to show the \textit{deviations} of these quantities from the $\Lambda$CDM predictions, e.g. $\Delta D^{ii}_\ell \equiv D^{ii}_\ell(f/|C_{\tau\ell}|) - D^{ii}_{\ell, \Lambda\mathrm{CDM}}$, where we defined
\begin{align}
    D_\ell^{TT} = \frac{\ell(\ell+1)C_\ell^{TT}}{2\pi}T_0^2, && D_\ell^{dd} = \frac{\ell^2 (\ell+1)^{2}}{2\pi}C_\ell^{\phi\phi} ,
\end{align}
with $T_0=2.7255$ K being the present-day CMB temperature~\cite{ParticleDataGroup:2024cfk}.
For our demonstration we take models with axion masses $m_a=\{0.1, 1, 10 \}$ eV and various values of $f/|C_{\tau l}|$ ranging from $10^5$ GeV to $10^9$ GeV and calculate the observables using \CLASS. All other cosmological parameters in these cases are set to their maximum-likelihood values derived from our MCMC scans. These predictions are compared to the Planck 2018 data \cite{Planck:2019nip, Planck:2018lbu}.

\begin{figure}[h!]
    \centering
    \includegraphics{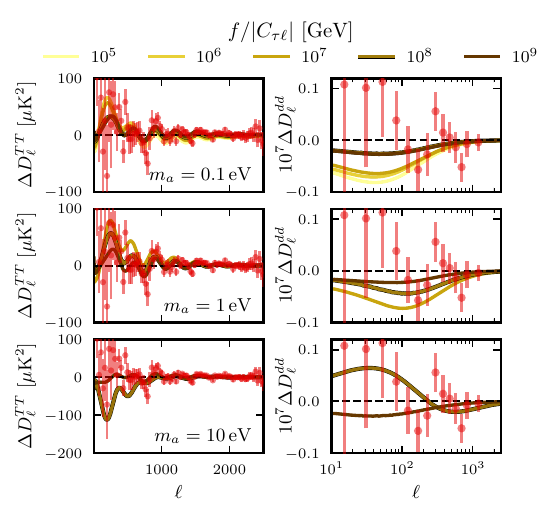}
    \caption{Differences in the CMB temperature power spectrum (left) and CMB lensing power spectrum (right) between the standard $\Lambda$CDM cosmology and scenarios with axion production via LFV $\tau$ couplings shown for axion masses $m_a = 0.1,\ 1,\ 10~\mathrm{eV}$ (from top to bottom). Each panel includes several representative values of $f/|C_{\tau l}|$ consistent with our parameter scans (cf. with Fig.~\ref{fig:lfv_mass_limits_scatter_plot}). Note that as the mass increases, fewer values of $f/|C_{\tau l}|$ are allowed. For the sake of comparison, the case $f/|C_{\tau l}| = 10^8~\mathrm{GeV}$ is highlighted on all panels. Planck 2018 data points are overlaid for reference. }
    \label{fig:PP_diff}
\end{figure}

In Fig.~\ref{fig:PP_diff} we show $\Delta D^{TT}_\ell$ and $\Delta D^{\phi\phi}_\ell$ predictions for different $f/|C_{\tau l}|$ modes from our scans and for the three aforementioned values of the axion mass. In case of $m_a=0.1$ eV all the lines corresponding to the values of $f/|C_{\tau l}|$ shown in the legend are present, whereas for $m_a=1$ and $10$ eV only the three and two highest values appear respectively. This is because the lower values of $f/|C_{\tau l}|$ cannot provide a satisfactory fit of the cosmological data used in our scan at higher axion masses, hence these scenarios are excluded. 
We highlight the curves corresponding to $f/|C_{\tau l}| = 10^8$ GeV with a black outline to trace how the constraints change with mass for a given coupling value. 

\bibliographystyle{elsarticle-num} 
\bibliography{references}

@article{Badziak:2024qjg,
    author = "Badziak, Marcin and Laletin, Maxim",
    title = "{Precise predictions for the QCD axion contribution to dark radiation with full phase-space evolution}",
    eprint = "2410.18186",
    archivePrefix = "arXiv",
    primaryClass = "hep-ph",
    doi = "10.1007/JHEP02(2025)108",
    journal = "JHEP",
    volume = "02",
    pages = "108",
    year = "2025"
}

@article{DEramo:2024jhn,
    author = "D'Eramo, Francesco and Lenoci, Alessandro",
    title = "{Back to the phase space: Thermal axion dark radiation via couplings to standard model fermions}",
    eprint = "2410.21253",
    archivePrefix = "arXiv",
    primaryClass = "hep-ph",
    doi = "10.1103/PhysRevD.110.116028",
    journal = "Phys. Rev. D",
    volume = "110",
    number = "11",
    pages = "116028",
    year = "2024"
}

@article{Notari:2022ffe,
	author = "Notari, Alessio and Rompineve, Fabrizio and Villadoro, Giovanni",
	title = "{Improved Hot Dark Matter Bound on the QCD Axion}",
	eprint = "2211.03799",
	archivePrefix = "arXiv",
	primaryClass = "hep-ph",
	reportNumber = "CERN-TH-2022-165",
	doi = "10.1103/PhysRevLett.131.011004",
	journal = "Phys. Rev. Lett.",
	volume = "131",
	number = "1",
	pages = "011004",
	year = "2023"
}

@article{GrillidiCortona:2015jxo,
	archiveprefix = {arXiv},
	author = {Grilli di Cortona, Giovanni and Hardy, Edward and Pardo Vega, Javier and Villadoro, Giovanni},
	doi = {10.1007/JHEP01(2016)034},
	eprint = {1511.02867},
	journal = {JHEP},
	pages = {034},
	primaryclass = {hep-ph},
	title = {{The QCD axion, precisely}},
	volume = {01},
	year = {2016},
}

@article{Boyarsky:2008xj,
    author = "Boyarsky, Alexey and Lesgourgues, Julien and Ruchayskiy, Oleg and Viel, Matteo",
    title = "{Lyman-alpha constraints on warm and on warm-plus-cold dark matter models}",
    eprint = "0812.0010",
    archivePrefix = "arXiv",
    primaryClass = "astro-ph",
    reportNumber = "CERN-PH-TH-2008-234, LAPTH-1290-08",
    doi = "10.1088/1475-7516/2009/05/012",
    journal = "JCAP",
    volume = "05",
    pages = "012",
    year = "2009"
}

@article{Baur:2017stq,
    author = "Baur, Julien and Palanque-Delabrouille, Nathalie and Yeche, Christophe and Boyarsky, Alexey and Ruchayskiy, Oleg and Armengaud, {\'E}ric and Lesgourgues, Julien",
    title = "{Constraints from Ly-$\alpha$ forests on non-thermal dark matter including resonantly-produced sterile neutrinos}",
    eprint = "1706.03118",
    archivePrefix = "arXiv",
    primaryClass = "astro-ph.CO",
    doi = "10.1088/1475-7516/2017/12/013",
    journal = "JCAP",
    volume = "12",
    pages = "013",
    year = "2017"
}

@article{Garcia-Gallego:2025kiw,
    author = "Garcia-Gallego, Olga and Ir{\v{s}}i{\v{c}}, Vid and Haehnelt, Martin G. and Viel, Matteo and Bolton, James S.",
    title = "{Constraining mixed dark matter models with high-redshift Lyman-alpha forest data}",
    eprint = "2504.06367",
    archivePrefix = "arXiv",
    primaryClass = "astro-ph.CO",
    doi = "10.1103/4k29-h99l",
    journal = "Phys. Rev. D",
    volume = "112",
    number = "4",
    pages = "043502",
    year = "2025"
}

@article{Ballesteros:2020adh,
    author = "Ballesteros, Guillermo and Garcia, Marcos A. G. and Pierre, Mathias",
    title = "{How warm are non-thermal relics? Lyman-$\alpha$ bounds on out-of-equilibrium dark matter}",
    eprint = "2011.13458",
    archivePrefix = "arXiv",
    primaryClass = "hep-ph",
    reportNumber = "IFT-UAM/CSIC-20-135",
    doi = "10.1088/1475-7516/2021/03/101",
    journal = "JCAP",
    volume = "03",
    pages = "101",
    year = "2021"
}

@article{DEramo:2020gpr,
    author = "D'Eramo, Francesco and Lenoci, Alessandro",
    title = "{Lower mass bounds on FIMP dark matter produced via freeze-in}",
    eprint = "2012.01446",
    archivePrefix = "arXiv",
    primaryClass = "hep-ph",
    reportNumber = "DESY 20-219, DESY-20-219",
    doi = "10.1088/1475-7516/2021/10/045",
    journal = "JCAP",
    volume = "10",
    pages = "045",
    year = "2021"
}

@article{Decant:2021mhj,
    author = "Decant, Quentin and Heisig, Jan and Hooper, Deanna C. and Lopez-Honorez, Laura",
    title = "{Lyman-{\ensuremath{\alpha}} constraints on freeze-in and superWIMPs}",
    eprint = "2111.09321",
    archivePrefix = "arXiv",
    primaryClass = "astro-ph.CO",
    reportNumber = "ULB-TH/21-20, TTK-21-46, HIP-2021-38/TH",
    doi = "10.1088/1475-7516/2022/03/041",
    journal = "JCAP",
    volume = "03",
    pages = "041",
    year = "2022"
}

@article{Kamada:2019kpe,
    author = "Kamada, Ayuki and Yanagi, Keisuke",
    title = "{Constraining FIMP from the structure formation of the Universe: analytic mapping from $m_{WDM}$}",
    eprint = "1907.04558",
    archivePrefix = "arXiv",
    primaryClass = "hep-ph",
    reportNumber = "CTPU-PTC-19-21, UT-19-16",
    doi = "10.1088/1475-7516/2019/11/029",
    journal = "JCAP",
    volume = "11",
    pages = "029",
    year = "2019"
}

@article{DEramo:2025jsb,
    author = "D'Eramo, Francesco and Lenoci, Alessandro and Dekker, Ariane",
    title = "{Dark Matter Freeze-In and Small-Scale Observables: Novel Mass Bounds and Viable Particle Candidates}",
    eprint = "2506.13864",
    archivePrefix = "arXiv",
    primaryClass = "hep-ph",
    month = "6",
    year = "2025"
}

@article{ParticleDataGroup:2024cfk,
    author = "Navas, S. and others",
    collaboration = "Particle Data Group",
    title = "{Review of particle physics}",
    doi = "10.1103/PhysRevD.110.030001",
    journal = "Phys. Rev. D",
    volume = "110",
    number = "3",
    pages = "030001",
    year = "2024"
}

@article{DEramo:2021lgb,
    author = "D'Eramo, Francesco and Hajkarim, Fazlollah and Yun, Seokhoon",
    title = "{Thermal QCD Axions across Thresholds}",
    eprint = "2108.05371",
    archivePrefix = "arXiv",
    primaryClass = "hep-ph",
    doi = "10.1007/JHEP10(2021)224",
    journal = "JHEP",
    volume = "10",
    pages = "224",
    year = "2021"
}

@article{DeSalas:2018rby,
    author = "De Salas, P. F. and Gariazzo, S. and Mena, O. and Ternes, C. A. and T{\'o}rtola, M.",
    title = "{Neutrino Mass Ordering from Oscillations and Beyond: 2018 Status and Future Prospects}",
    eprint = "1806.11051",
    archivePrefix = "arXiv",
    primaryClass = "hep-ph",
    doi = "10.3389/fspas.2018.00036",
    journal = "Front. Astron. Space Sci.",
    volume = "5",
    pages = "36",
    year = "2018"
}

@article{CLASS2,
    author = "Blas, Diego and Lesgourgues, Julien and Tram, Thomas",
    title = "{The Cosmic Linear Anisotropy Solving System (CLASS) II: Approximation schemes}",
    eprint = "1104.2933",
    archivePrefix = "arXiv",
    primaryClass = "astro-ph.CO",
    reportNumber = "CERN-PH-TH-2011-082, LAPTH-010-11",
    doi = "10.1088/1475-7516/2011/07/034",
    journal = "JCAP",
    volume = "07",
    pages = "034",
    year = "2011"
}

@article{CLASS4,
    author = "Lesgourgues, Julien and Tram, Thomas",
    title = "{The Cosmic Linear Anisotropy Solving System (CLASS) IV: efficient implementation of non-cold relics}",
    eprint = "1104.2935",
    archivePrefix = "arXiv",
    primaryClass = "astro-ph.CO",
    reportNumber = "CERN-PH-TH-2011-084, LAPTH-012-11",
    doi = "10.1088/1475-7516/2011/09/032",
    journal = "JCAP",
    volume = "09",
    pages = "032",
    year = "2011"
}

@article{Brinckmann:2018cvx,
      author         = "Brinckmann, Thejs and Lesgourgues, Julien",
      title          = "{MontePython 3: boosted MCMC sampler and other features}",
      year           = "2018",
      eprint         = "1804.07261",
      archivePrefix  = "arXiv",
      primaryClass   = "astro-ph.CO",
}

@article{Audren:2012wb,
      author         = "Audren, Benjamin and Lesgourgues, Julien and Benabed,
                        Karim and Prunet, Simon",
      title          = "{Conservative Constraints on Early Cosmology: an
                        illustration of the Monte Python cosmological parameter
                        inference code}",
      journal        = "JCAP",
      volume         = "1302",
      pages          = "001",
      doi            = "10.1088/1475-7516/2013/02/001",
      year           = "2013",
      eprint         = "1210.7183",
      archivePrefix  = "arXiv",
      primaryClass   = "astro-ph.CO",
      reportNumber   = "CERN-PH-TH-2012-290, LAPTH-048-12",
}

@article{Lewis:2019xzd,
 author         = "Lewis, Antony",
 title          = "{GetDist: a Python package for analysing Monte Carlo
                   samples}",
 year           = "2019",
 eprint         = "1910.13970",
 archivePrefix  = "arXiv",
 primaryClass   = "astro-ph.IM",

 url            = "https://getdist.readthedocs.io"
}

@article{Planck:2018vyg,
    author = "Aghanim, N. and others",
    collaboration = "Planck",
    title = "{Planck 2018 results. VI. Cosmological parameters}",
    eprint = "1807.06209",
    archivePrefix = "arXiv",
    primaryClass = "astro-ph.CO",
    doi = "10.1051/0004-6361/201833910",
    journal = "Astron. Astrophys.",
    volume = "641",
    pages = "A6",
    year = "2020",
    note = "[Erratum: Astron.Astrophys. 652, C4 (2021)]"
}

@article{Planck:2019nip,
    author = "Aghanim, N. and others",
    collaboration = "Planck",
    title = "{Planck 2018 results. V. CMB power spectra and likelihoods}",
    eprint = "1907.12875",
    archivePrefix = "arXiv",
    primaryClass = "astro-ph.CO",
    doi = "10.1051/0004-6361/201936386",
    journal = "Astron. Astrophys.",
    volume = "641",
    pages = "A5",
    year = "2020"
}

@article{Planck:2018lbu,
    author = "Aghanim, N. and others",
    collaboration = "Planck",
    title = "{Planck 2018 results. VIII. Gravitational lensing}",
    eprint = "1807.06210",
    archivePrefix = "arXiv",
    primaryClass = "astro-ph.CO",
    doi = "10.1051/0004-6361/201833886",
    journal = "Astron. Astrophys.",
    volume = "641",
    pages = "A8",
    year = "2020"
}

@article{DESI:2025zgx,
    author = "Abdul Karim, M. and others",
    collaboration = "DESI",
    title = "{DESI DR2 results. II. Measurements of baryon acoustic oscillations and cosmological constraints}",
    eprint = "2503.14738",
    archivePrefix = "arXiv",
    primaryClass = "astro-ph.CO",
    reportNumber = "FERMILAB-PUB-25-0169-PPD",
    doi = "10.1103/tr6y-kpc6",
    journal = "Phys. Rev. D",
    volume = "112",
    number = "8",
    pages = "083515",
    year = "2025"
}

@article{Aghaie:2024jkj,
    author = "Aghaie, Mohammad and Armando, Giovanni and Conaci, Angela and Dondarini, Alessandro and Matak, Peter and Panci, Paolo and Sinska, Zuzana and Ziegler, Robert",
    title = "{Axion dark matter from heavy quarks}",
    eprint = "2404.12199",
    archivePrefix = "arXiv",
    primaryClass = "hep-ph",
    doi = "10.1016/j.physletb.2024.138923",
    journal = "Phys. Lett. B",
    volume = "856",
    pages = "138923",
    year = "2024"
}

@article{Panci:2022wlc,
    author = "Panci, Paolo and Redigolo, Diego and Schwetz, Thomas and Ziegler, Robert",
    title = "{Axion dark matter from lepton flavor-violating decays}",
    eprint = "2209.03371",
    archivePrefix = "arXiv",
    primaryClass = "hep-ph",
    doi = "10.1016/j.physletb.2023.137919",
    journal = "Phys. Lett. B",
    volume = "841",
    pages = "137919",
    year = "2023"
}

@article{DEramo:2022nvb,
	archiveprefix = {arXiv},
	author = {D'Eramo, Francesco and Di Valentino, Eleonora and Giar\`e, William and Hajkarim, Fazlollah and Melchiorri, Alessandro and Mena, Olga and Renzi, Fabrizio and Yun, Seokhoon},
	doi = {10.1088/1475-7516/2022/09/022},
	eprint = {2205.07849},
	journal = {JCAP},
	pages = {022},
	primaryclass = {astro-ph.CO},
	title = {{Cosmological bound on the QCD axion mass, redux}},
	volume = {09},
	year = {2022},
}

@article{Binder:2017rgn,
	author = "Binder, Tobias and Bringmann, Torsten and Gustafsson, Michael and Hryczuk, Andrzej",
	title = "{Early kinetic decoupling of dark matter: when the standard way of calculating the thermal relic density fails}",
	eprint = "1706.07433",
	archivePrefix = "arXiv",
	primaryClass = "astro-ph.CO",
	doi = "10.1103/PhysRevD.96.115010",
	journal = "Phys. Rev. D",
	volume = "96",
	number = "11",
	pages = "115010",
	year = "2017",
	note = "[Erratum: Phys.Rev.D 101, 099901 (2020)]"
}

@article{DEramo:2018vss,
	author = "D'Eramo, Francesco and Ferreira, Ricardo Z. and Notari, Alessio and Bernal, Jos\'e Luis",
	title = "{Hot Axions and the $H_0$ tension}",
	eprint = "1808.07430",
	archivePrefix = "arXiv",
	primaryClass = "hep-ph",
	doi = "10.1088/1475-7516/2018/11/014",
	journal = "JCAP",
	volume = "11",
	pages = "014",
	year = "2018"
}

@article{Green:2021hjh,
	author = "Green, Daniel and Guo, Yi and Wallisch, Benjamin",
	title = "{Cosmological implications of axion-matter couplings}",
	eprint = "2109.12088",
	archivePrefix = "arXiv",
	primaryClass = "astro-ph.CO",
	doi = "10.1088/1475-7516/2022/02/019",
	journal = "JCAP",
	volume = "02",
	number = "02",
	pages = "019",
	year = "2022"
}

@article{Caloni:2022uya,
	archiveprefix = {arXiv},
	author = {Caloni, Luca and Gerbino, Martina and Lattanzi, Massimiliano and Visinelli, Luca},
	doi = {10.1088/1475-7516/2022/09/021},
	eprint = {2205.01637},
	journal = {JCAP},
	pages = {021},
	primaryclass = {astro-ph.CO},
	title = {{Novel cosmological bounds on thermally-produced axion-like particles}},
	volume = {09},
	year = {2022}}

@article{TWIST:2014ymv,
	archiveprefix = {arXiv},
	author = {Bayes, R. and others},
	collaboration = {TWIST},
	doi = {10.1103/PhysRevD.91.052020},
	eprint = {1409.0638},
	journal = {Phys. Rev. D},
	number = {5},
	pages = {052020},
	primaryclass = {hep-ex},
	title = {{Search for two body muon decay signals}},
	volume = {91},
	year = {2015}}

@article{Calibbi:2020jvd,
	archiveprefix = {arXiv},
	author = {Calibbi, Lorenzo and Redigolo, Diego and Ziegler, Robert and Zupan, Jure},
	doi = {10.1007/JHEP09(2021)173},
	eprint = {2006.04795},
	journal = {JHEP},
	pages = {173},
	primaryclass = {hep-ph},
	reportnumber = {P3H-20-024, TTP20-025},
	title = {{Looking forward to lepton-flavor-violating ALPs}},
	volume = {09},
	year = {2021}}

@article{Ferreira:2020bpb,
    author = "Ferreira, Ricardo Z. and Notari, Alessio and Rompineve, Fabrizio",
    title = "{Dine-Fischler-Srednicki-Zhitnitsky axion in the CMB}",
    eprint = "2012.06566",
    archivePrefix = "arXiv",
    primaryClass = "hep-ph",
    doi = "10.1103/PhysRevD.103.063524",
    journal = "Phys. Rev. D",
    volume = "103",
    number = "6",
    pages = "063524",
    year = "2021"
}

@article{Badziak:2024szg,
    author = "Badziak, Marcin and Harigaya, Keisuke and \L{}ukawski, Micha\l{} and Ziegler, Robert",
    title = "{Thermal production of astrophobic axions}",
    eprint = "2403.05621",
    archivePrefix = "arXiv",
    primaryClass = "hep-ph",
    doi = "10.1007/JHEP09(2024)136",
    journal = "JHEP",
    volume = "09",
    pages = "136",
    year = "2024"
}

@article{Peccei:1977hh,
	author = {Peccei, R. D. and Quinn, Helen R.},
	doi = {10.1103/PhysRevLett.38.1440},
	journal = {Phys. Rev. Lett.},
	pages = {1440--1443},
	reportnumber = {ITP-568-STANFORD},
	title = {{CP Conservation in the Presence of Instantons}},
	volume = {38},
	year = {1977},
}

@article{Preskill:1982cy,
    author = "Preskill, John and Wise, Mark B. and Wilczek, Frank",
    editor = "Srednicki, M. A.",
    title = "{Cosmology of the Invisible Axion}",
    reportNumber = "HUTP-82-A048, NSF-ITP-82-103",
    doi = "10.1016/0370-2693(83)90637-8",
    journal = "Phys. Lett. B",
    volume = "120",
    pages = "127--132",
    year = "1983"
}

@article{Abbott:1982af,
    author = "Abbott, L. F. and Sikivie, P.",
    editor = "Srednicki, M. A.",
    title = "{A Cosmological Bound on the Invisible Axion}",
    reportNumber = "PRINT-82-0695 (BRANDEIS)",
    doi = "10.1016/0370-2693(83)90638-X",
    journal = "Phys. Lett. B",
    volume = "120",
    pages = "133--136",
    year = "1983"
}

@article{Dine:1982ah,
    author = "Dine, Michael and Fischler, Willy",
    editor = "Srednicki, M. A.",
    title = "{The Not So Harmless Axion}",
    reportNumber = "UPR-0201T",
    doi = "10.1016/0370-2693(83)90639-1",
    journal = "Phys. Lett. B",
    volume = "120",
    pages = "137--141",
    year = "1983"
}

@article{Kim:1979if,
	author = {Kim, Jihn E.},
	doi = {10.1103/PhysRevLett.43.103},
	journal = {Phys. Rev. Lett.},
	pages = {103},
	reportnumber = {UPR-0120T},
	title = {{Weak Interaction Singlet and Strong CP Invariance}},
	volume = {43},
	year = {1979},
}

@article{Shifman:1979if,
	author = {Shifman, Mikhail A. and Vainshtein, A. I. and Zakharov, Valentin I.},
	doi = {10.1016/0550-3213(80)90209-6},
	journal = {Nucl. Phys. B},
	pages = {493--506},
	reportnumber = {ITEP-64-1979},
	title = {{Can Confinement Ensure Natural CP Invariance of Strong Interactions?}},
	volume = {166},
	year = {1980},
}

@article{Dine:1981rt,
	author = {Dine, Michael and Fischler, Willy and Srednicki, Mark},
	doi = {10.1016/0370-2693(81)90590-6},
	journal = {Phys. Lett. B},
	pages = {199--202},
	reportnumber = {Print-81-0320 (IAS,PRINCETON)},
	title = {{A Simple Solution to the Strong CP Problem with a Harmless Axion}},
	volume = {104},
	year = {1981},
}

@article{Zhitnitsky:1980tq,
	author = {Zhitnitsky, A. R.},
	journal = {Sov. J. Nucl. Phys.},
	pages = {260},
	title = {{On Possible Suppression of the Axion Hadron Interactions. (In Russian)}},
	volume = {31},
	year = {1980}}

@article{Buschmann:2021juv,
	archiveprefix = {arXiv},
	author = {Buschmann, Malte and Dessert, Christopher and Foster, Joshua W. and Long, Andrew J. and Safdi, Benjamin R.},
	doi = {10.1103/PhysRevLett.128.091102},
	eprint = {2111.09892},
	journal = {Phys. Rev. Lett.},
	number = {9},
	pages = {091102},
	primaryclass = {hep-ph},
	title = {{Upper Limit on the QCD Axion Mass from Isolated Neutron Star Cooling}},
	volume = {128},
	year = {2022},
}

@article{DiLuzio:2017ogq,
	archiveprefix = {arXiv},
	author = {Di Luzio, Luca and Mescia, Federico and Nardi, Enrico and Panci, Paolo and Ziegler, Robert},
	doi = {10.1103/PhysRevLett.120.261803},
	eprint = {1712.04940},
	journal = {Phys. Rev. Lett.},
	number = {26},
	pages = {261803},
	primaryclass = {hep-ph},
	reportnumber = {IPPP-17-102, CERN-TH-2017-256},
	title = {{Astrophobic Axions}},
	volume = {120},
	year = {2018},
}

@article{Badziak:2023fsc,
	archiveprefix = {arXiv},
	author = {Badziak, Marcin and Harigaya, Keisuke},
	doi = {10.1007/JHEP06(2023)014},
	eprint = {2301.09647},
	journal = {JHEP},
	pages = {014},
	primaryclass = {hep-ph},
	title = {{Naturally astrophobic QCD axion}},
	volume = {06},
	year = {2023},
}

@article{Badziak:2021apn,
	archiveprefix = {arXiv},
	author = {Badziak, Marcin and Grilli di Cortona, Giovanni and Tabet, Mustafa and Ziegler, Robert},
	doi = {10.1007/JHEP10(2021)181},
	eprint = {2107.09708},
	journal = {JHEP},
	pages = {181},
	primaryclass = {hep-ph},
	reportnumber = {TTP21-025, P3H-21-051},
	title = {{Flavor-violating Higgs decays and stellar cooling anomalies in axion models}},
	volume = {10},
	year = {2021},
}

@article{Bjorkeroth:2019jtx,
	archiveprefix = {arXiv},
	author = {Bj\"orkeroth, Fredrik and Di Luzio, Luca and Mescia, Federico and Nardi, Enrico and Panci, Paolo and Ziegler, Robert},
	doi = {10.1103/PhysRevD.101.035027},
	eprint = {1907.06575},
	journal = {Phys. Rev. D},
	number = {3},
	pages = {035027},
	primaryclass = {hep-ph},
	reportnumber = {CERN-TH-2019-118, DESY-19-194},
	title = {{Axion-electron decoupling in nucleophobic axion models}},
	volume = {101},
	year = {2020},
}

@article{Chang:1993gm,
	archiveprefix = {arXiv},
	author = {Chang, Sanghyeon and Choi, Kiwoon},
	doi = {10.1016/0370-2693(93)90656-3},
	eprint = {hep-ph/9306216},
	journal = {Phys. Lett. B},
	pages = {51--56},
	reportnumber = {SNUTP-93-11},
	title = {{Hadronic axion window and the big bang nucleosynthesis}},
	volume = {316},
	year = {1993},
}

@article{Hannestad:2005df,
	archiveprefix = {arXiv},
	author = {Hannestad, Steen and Mirizzi, Alessandro and Raffelt, Georg},
	doi = {10.1088/1475-7516/2005/07/002},
	eprint = {hep-ph/0504059},
	journal = {JCAP},
	pages = {002},
	title = {{New cosmological mass limit on thermal relic axions}},
	volume = {07},
	year = {2005},
}

@article{SimonsObservatory:2018koc,
    author = "Ade, Peter and others",
    collaboration = "Simons Observatory",
    title = "{The Simons Observatory: Science goals and forecasts}",
    eprint = "1808.07445",
    archivePrefix = "arXiv",
    primaryClass = "astro-ph.CO",
    doi = "10.1088/1475-7516/2019/02/056",
    journal = "JCAP",
    volume = "02",
    pages = "056",
    year = "2019"
}

@article{DEramo:2023nzt,
    author = "D'Eramo, Francesco and Hajkarim, Fazlollah and Lenoci, Alessandro",
    title = "{Dark radiation from the primordial thermal bath in momentum space}",
    eprint = "2311.04974",
    archivePrefix = "arXiv",
    primaryClass = "hep-ph",
    reportNumber = "DESY-23-177",
    doi = "10.1088/1475-7516/2024/03/009",
    journal = "JCAP",
    volume = "03",
    pages = "009",
    year = "2024"
}

@article{Hall:2009bx,
    author = "Hall, Lawrence J. and Jedamzik, Karsten and March-Russell, John and West, Stephen M.",
    title = "{Freeze-In Production of FIMP Dark Matter}",
    eprint = "0911.1120",
    archivePrefix = "arXiv",
    primaryClass = "hep-ph",
    reportNumber = "OUTP-09-18-P, UCB-PTH-09-32",
    doi = "10.1007/JHEP03(2010)080",
    journal = "JHEP",
    volume = "03",
    pages = "080",
    year = "2010"
}

@article{DEramo:2021usm,
    author = "D'Eramo, Francesco and Yun, Seokhoon",
    title = "{Flavor violating axions in the early Universe}",
    eprint = "2111.12108",
    archivePrefix = "arXiv",
    primaryClass = "hep-ph",
    doi = "10.1103/PhysRevD.105.075002",
    journal = "Phys. Rev. D",
    volume = "105",
    number = "7",
    pages = "075002",
    year = "2022"
}

@article{Ferreira:2018vjj,
	archiveprefix = {arXiv},
	author = {Ferreira, Ricardo Z. and Notari, Alessio},
	doi = {10.1103/PhysRevLett.120.191301},
	eprint = {1801.06090},
	journal = {Phys. Rev. Lett.},
	number = {19},
	pages = {191301},
	primaryclass = {hep-ph},
	title = {{Observable Windows for the QCD Axion Through the Number of Relativistic Species}},
	volume = {120},
	year = {2018},
}

@article{Arias-Aragon:2020shv,
	archiveprefix = {arXiv},
	author = {Arias-Arag\'on, Fernando and D'Eramo, Francesco and Ferreira, Ricardo Z. and Merlo, Luca and Notari, Alessio},
	doi = {10.1088/1475-7516/2021/03/090},
	eprint = {2012.04736},
	journal = {JCAP},
	pages = {090},
	primaryclass = {hep-ph},
	title = {{Production of Thermal Axions across the ElectroWeak Phase Transition}},
	volume = {03},
	year = {2021},
}

@article{Salvio:2013iaa,
	archiveprefix = {arXiv},
	author = {Salvio, Alberto and Strumia, Alessandro and Xue, Wei},
	doi = {10.1088/1475-7516/2014/01/011},
	eprint = {1310.6982},
	journal = {JCAP},
	pages = {011},
	primaryclass = {hep-ph},
	reportnumber = {FTUAM-13-29, IFT-UAM-CSIC-13-113},
	title = {{Thermal axion production}},
	volume = {01},
	year = {2014},
}

@article{DEramo:2021psx,
    author = "D'Eramo, Francesco and Hajkarim, Fazlollah and Yun, Seokhoon",
    title = "{Thermal Axion Production at Low Temperatures: A Smooth Treatment of the QCD Phase Transition}",
    eprint = "2108.04259",
    archivePrefix = "arXiv",
    primaryClass = "hep-ph",
    doi = "10.1103/PhysRevLett.128.152001",
    journal = "Phys. Rev. Lett.",
    volume = "128",
    number = "15",
    pages = "152001",
    year = "2022"
}

@article{Calibbi:2016hwq,
    author = "Calibbi, Lorenzo and Goertz, Florian and Redigolo, Diego and Ziegler, Robert and Zupan, Jure",
    title = "{Minimal axion model from flavor}",
    eprint = "1612.08040",
    archivePrefix = "arXiv",
    primaryClass = "hep-ph",
    reportNumber = "TTP16-058, CERN-TH-2016-261",
    doi = "10.1103/PhysRevD.95.095009",
    journal = "Phys. Rev. D",
    volume = "95",
    number = "9",
    pages = "095009",
    year = "2017"
}

@article{Ema:2016ops,
    author = "Ema, Yohei and Hamaguchi, Koichi and Moroi, Takeo and Nakayama, Kazunori",
    title = "{Flaxion: a minimal extension to solve puzzles in the standard model}",
    eprint = "1612.05492",
    archivePrefix = "arXiv",
    primaryClass = "hep-ph",
    reportNumber = "UT-16-36, IPMU16-0189",
    doi = "10.1007/JHEP01(2017)096",
    journal = "JHEP",
    volume = "01",
    pages = "096",
    year = "2017"
}

@article{Belle-II:2022heu,
	archiveprefix = {arXiv},
	author = {Adachi, I. and others},
	collaboration = {Belle-II},
	doi = {10.1103/PhysRevLett.130.181803},
	eprint = {2212.03634},
	journal = {Phys. Rev. Lett.},
	number = {18},
	pages = {181803},
	primaryclass = {hep-ex},
	reportnumber = {Belle II Preprint 2022-007, KEK Preprint 2022-39},
	title = {{Search for Lepton-Flavor-Violating \ensuremath{\tau} Decays to a Lepton and an Invisible Boson at Belle II}},
	volume = {130},
	year = {2023},
}

@article{Dunsky:2022uoq,
    author = "Dunsky, David I. and Hall, Lawrence J. and Harigaya, Keisuke",
    title = "{Dark Radiation Constraints on Heavy QCD Axions}",
    eprint = "2205.11540",
    archivePrefix = "arXiv",
    primaryClass = "hep-ph",
    doi = "10.1007/JHEP04(2024)130",
    journal = "JHEP",
    volume = "04",
    pages = "130",
    year = "2024"
}

@article{Bianchini:2023ubu,
    author = "Bianchini, Federico and di Cortona, Giovanni Grilli and Valli, Mauro",
    title = "{QCD axion: Some like it hot}",
    eprint = "2310.08169",
    archivePrefix = "arXiv",
    primaryClass = "hep-ph",
    doi = "10.1103/PhysRevD.110.123527",
    journal = "Phys. Rev. D",
    volume = "110",
    number = "12",
    pages = "123527",
    year = "2024"
}

@article{Bouzoud:2024bom,
    author = "Bouzoud, Killian and Ghiglieri, Jacopo",
    title = "{Thermal axion production at hard and soft momenta}",
    eprint = "2404.06113",
    archivePrefix = "arXiv",
    primaryClass = "hep-ph",
    doi = "10.1007/JHEP01(2025)163",
    journal = "JHEP",
    volume = "01",
    pages = "163",
    year = "2025"
}

@article{Carenza:2020cis,
    author = "Carenza, Pierluca and Fore, Bryce and Giannotti, Maurizio and Mirizzi, Alessandro and Reddy, Sanjay",
    title = "{Enhanced Supernova Axion Emission and its Implications}",
    eprint = "2010.02943",
    archivePrefix = "arXiv",
    primaryClass = "hep-ph",
    reportNumber = "INT-PUB-20-039",
    doi = "10.1103/PhysRevLett.126.071102",
    journal = "Phys. Rev. Lett.",
    volume = "126",
    number = "7",
    pages = "071102",
    year = "2021"
}

@article{MillerBertolami:2014rka,
    author = "Miller Bertolami, Marcelo M. and Melendez, Brenda E. and Althaus, Leandro G. and Isern, Jordi",
    title = "{Revisiting the axion bounds from the Galactic white dwarf luminosity function}",
    eprint = "1406.7712",
    archivePrefix = "arXiv",
    primaryClass = "hep-ph",
    doi = "10.1088/1475-7516/2014/10/069",
    journal = "JCAP",
    volume = "10",
    pages = "069",
    year = "2014"
}

@article{Caputo:2021rux,
    author = "Caputo, Andrea and Raffelt, Georg and Vitagliano, Edoardo",
    title = "{Muonic boson limits: Supernova redux}",
    eprint = "2109.03244",
    archivePrefix = "arXiv",
    primaryClass = "hep-ph",
    reportNumber = "MPP-2021-154",
    doi = "10.1103/PhysRevD.105.035022",
    journal = "Phys. Rev. D",
    volume = "105",
    number = "3",
    pages = "035022",
    year = "2022"
}

@article{PhysRevD.34.1967,
  title = {Search for right-handed currents in muon decay},
  author = {Jodidio, A. and Balke, B. and Carr, J. and Gidal, G. and Shinsky, K. A. and Steiner, H. M. and Stoker, D. P. and Strovink, M. and Tripp, R. D. and Gobbi, B. and Oram, C. J.},
  journal = {Phys. Rev. D},
  volume = {34},
  issue = {7},
  pages = {1967--1990},
  numpages = {0},
  year = {1986},
  month = {Oct},
  publisher = {American Physical Society},
  doi = {10.1103/PhysRevD.34.1967},
  url = {https://link.aps.org/doi/10.1103/PhysRevD.34.1967}
}

@article{Co:2019wyp,
    author = "Co, Raymond T. and Harigaya, Keisuke",
    title = "{Axiogenesis}",
    eprint = "1910.02080",
    archivePrefix = "arXiv",
    primaryClass = "hep-ph",
    reportNumber = "LCTP-19-27",
    doi = "10.1103/PhysRevLett.124.111602",
    journal = "Phys. Rev. Lett.",
    volume = "124",
    number = "11",
    pages = "111602",
    year = "2020"
}

@article{Takahashi:2023vhv,
    author = "Takahashi, Fuminobu and Yin, Wen",
    title = "{Hadrophobic axion from a GUT}",
    eprint = "2301.10757",
    archivePrefix = "arXiv",
    primaryClass = "hep-ph",
    reportNumber = "TU-1179",
    doi = "10.1103/PhysRevD.109.035024",
    journal = "Phys. Rev. D",
    volume = "109",
    number = "3",
    pages = "035024",
    year = "2024"
}

@article{DiLuzio:2022tyc,
    author = "Di Luzio, Luca and Mescia, Federico and Nardi, Enrico and Okawa, Shohei",
    title = "{Renormalization group effects in astrophobic axion models}",
    eprint = "2205.15326",
    archivePrefix = "arXiv",
    primaryClass = "hep-ph",
    doi = "10.1103/PhysRevD.106.055016",
    journal = "Phys. Rev. D",
    volume = "106",
    number = "5",
    pages = "055016",
    year = "2022"
}

@article{Ghosh:2020vti,
    author = "Ghosh, Diptimoy and Sachdeva, Divya",
    title = "{Constraints on Axion-Lepton coupling from Big Bang Nucleosynthesis}",
    eprint = "2007.01873",
    archivePrefix = "arXiv",
    primaryClass = "hep-ph",
    doi = "10.1088/1475-7516/2020/10/060",
    journal = "JCAP",
    volume = "10",
    pages = "060",
    year = "2020"
}

@article{Langhoff:2022bij,
    author = "Langhoff, Kevin and Outmezguine, Nadav Joseph and Rodd, Nicholas L.",
    title = "{Irreducible Axion Background}",
    eprint = "2209.06216",
    archivePrefix = "arXiv",
    primaryClass = "hep-ph",
    reportNumber = "CERN-TH-2022-148",
    doi = "10.1103/PhysRevLett.129.241101",
    journal = "Phys. Rev. Lett.",
    volume = "129",
    number = "24",
    pages = "241101",
    year = "2022"
}

@article{Vilenkin:1982ks,
    author = "Vilenkin, A. and Everett, A. E.",
    title = "{Cosmic Strings and Domain Walls in Models with Goldstone and PseudoGoldstone Bosons}",
    doi = "10.1103/PhysRevLett.48.1867",
    journal = "Phys. Rev. Lett.",
    volume = "48",
    pages = "1867--1870",
    year = "1982"
}

@article{Kawasaki:2014sqa,
    author = "Kawasaki, Masahiro and Saikawa, Ken'ichi and Sekiguchi, Toyokazu",
    title = "{Axion dark matter from topological defects}",
    eprint = "1412.0789",
    archivePrefix = "arXiv",
    primaryClass = "hep-ph",
    reportNumber = "ICRR-REPORT-696-2014-22, IPMU14-0348",
    doi = "10.1103/PhysRevD.91.065014",
    journal = "Phys. Rev. D",
    volume = "91",
    number = "6",
    pages = "065014",
    year = "2015"
}

@article{Buschmann:2019icd,
    author = "Buschmann, Malte and Foster, Joshua W. and Safdi, Benjamin R.",
    title = "{Early-Universe Simulations of the Cosmological Axion}",
    eprint = "1906.00967",
    archivePrefix = "arXiv",
    primaryClass = "astro-ph.CO",
    reportNumber = "LCTP-19-08",
    doi = "10.1103/PhysRevLett.124.161103",
    journal = "Phys. Rev. Lett.",
    volume = "124",
    number = "16",
    pages = "161103",
    year = "2020"
}

@article{Gorghetto:2020qws,
    author = "Gorghetto, Marco and Hardy, Edward and Villadoro, Giovanni",
    title = "{More axions from strings}",
    eprint = "2007.04990",
    archivePrefix = "arXiv",
    primaryClass = "hep-ph",
    doi = "10.21468/SciPostPhys.10.2.050",
    journal = "SciPost Phys.",
    volume = "10",
    number = "2",
    pages = "050",
    year = "2021"
}

@article{Co:2019jts,
    author = "Co, Raymond T. and Hall, Lawrence J. and Harigaya, Keisuke",
    title = "{Axion Kinetic Misalignment Mechanism}",
    eprint = "1910.14152",
    archivePrefix = "arXiv",
    primaryClass = "hep-ph",
    reportNumber = "LCTP-19-28",
    doi = "10.1103/PhysRevLett.124.251802",
    journal = "Phys. Rev. Lett.",
    volume = "124",
    number = "25",
    pages = "251802",
    year = "2020"
}

@article{Allali:2024cji,
    author = "Allali, Itamar J. and Notari, Alessio and Rompineve, Fabrizio",
    title = "{Reduced Hubble tension in dark radiation models after DESI 2024}",
    eprint = "2404.15220",
    archivePrefix = "arXiv",
    primaryClass = "astro-ph.CO",
    doi = "10.1088/1475-7516/2025/03/023",
    journal = "JCAP",
    volume = "03",
    pages = "023",
    year = "2025"
}

@article{Xu:2021rwg,
    author = "Xu, Weishuang Linda and Mu{\~n}oz, Julian B. and Dvorkin, Cora",
    title = "{Cosmological constraints on light but massive relics}",
    eprint = "2107.09664",
    archivePrefix = "arXiv",
    primaryClass = "astro-ph.CO",
    doi = "10.1103/PhysRevD.105.095029",
    journal = "Phys. Rev. D",
    volume = "105",
    number = "9",
    pages = "095029",
    year = "2022"
}

@article{Heymans:2013fya,
    author = "Heymans, Catherine and others",
    title = "{CFHTLenS tomographic weak lensing cosmological parameter constraints: Mitigating the impact of intrinsic galaxy alignments}",
    eprint = "1303.1808",
    archivePrefix = "arXiv",
    primaryClass = "astro-ph.CO",
    doi = "10.1093/mnras/stt601",
    journal = "Mon. Not. Roy. Astron. Soc.",
    volume = "432",
    pages = "2433",
    year = "2013"
}

@article{Takahashi:2012em,
    author = "Takahashi, Ryuichi and Sato, Masanori and Nishimichi, Takahiro and Taruya, Atsushi and Oguri, Masamune",
    title = "{Revising the Halofit Model for the Nonlinear Matter Power Spectrum}",
    eprint = "1208.2701",
    archivePrefix = "arXiv",
    primaryClass = "astro-ph.CO",
    doi = "10.1088/0004-637X/761/2/152",
    journal = "Astrophys. J.",
    volume = "761",
    pages = "152",
    year = "2012"
}

@article{Ali-Haimoud:2012fzp,
    author = "Ali-Haimoud, Yacine and Bird, Simeon",
    title = "{An efficient implementation of massive neutrinos in non-linear structure formation simulations}",
    eprint = "1209.0461",
    archivePrefix = "arXiv",
    primaryClass = "astro-ph.CO",
    doi = "10.1093/mnras/sts286",
    journal = "Mon. Not. Roy. Astron. Soc.",
    volume = "428",
    pages = "3375--3389",
    year = "2012"
}

@article{Berezhiani:1992rk,
    author = "Berezhiani, Z. G. and Sakharov, A. S. and Khlopov, M. Yu.",
    title = "{Primordial background of cosmological axions}",
    journal = "Sov. J. Nucl. Phys.",
    volume = "55",
    pages = "1063--1071",
    year = "1992"
}

@article{Zhang:2023vva,
    author = "Zhang, Hong-Yi and Hagimoto, Ray and Long, Andrew J.",
    title = "{Neutron star cooling with lepton-flavor-violating axions}",
    eprint = "2309.03889",
    archivePrefix = "arXiv",
    primaryClass = "hep-ph",
    doi = "10.1103/PhysRevD.109.103005",
    journal = "Phys. Rev. D",
    volume = "109",
    number = "10",
    pages = "103005",
    year = "2024"
}

@article{Davidson:1981zd,
    author = "Davidson, Aharon and Wali, Kameshwar C.",
    title = "{MINIMAL FLAVOR UNIFICATION VIA MULTIGENERATIONAL PECCEI-QUINN SYMMETRY}",
    reportNumber = "WIS-81/40-Ph, SU-4217-206",
    doi = "10.1103/PhysRevLett.48.11",
    journal = "Phys. Rev. Lett.",
    volume = "48",
    pages = "11",
    year = "1982"
}

@article{Wilczek:1982rv,
    author = "Wilczek, Frank",
    title = "{Axions and Family Symmetry Breaking}",
    doi = "10.1103/PhysRevLett.49.1549",
    journal = "Phys. Rev. Lett.",
    volume = "49",
    pages = "1549--1552",
    year = "1982"
}

@article{Berezhiani:1989fp,
    author = "Berezhiani, Z. G. and Khlopov, M. Yu.",
    title = "{Cosmology of Spontaneously Broken Gauge Family Symmetry}",
    reportNumber = "FERMILAB-PUB-89-202-A",
    doi = "10.1007/BF01570798",
    journal = "Z. Phys. C",
    volume = "49",
    pages = "73--78",
    year = "1991"
}

@article{MartinCamalich:2025srw,
    author = "Martin Camalich, Jorge and Ziegler, Robert",
    title = "{Flavor Phenomenology of Light Dark Sectors}",
    eprint = "2503.17323",
    archivePrefix = "arXiv",
    primaryClass = "hep-ph",
    doi = "10.1146/annurev-nucl-121423-100931",
    journal = "Ann. Rev. Nucl. Part. Sci.",
    volume = "75",
    number = "1",
    pages = "223--246",
    year = "2025"
}

\end{document}